\documentclass[]{PoS}
\usepackage{natbib}
\usepackage{mathptmx}       
\usepackage{amssymb}
\usepackage{fontenc}
\usepackage{times}
\usepackage{newtxtext,newtxmath,amsmath}
\usepackage{graphicx}        

\title{Hydrodynamic Simulations of Classical Novae; CO and ONe White Dwarfs are Supernova Ia Progenitors}

\ShortTitle{Hydrodynamic Simulations of Classical Novae}

\author{\speaker{S. Starrfield} 
       \thanks{This work supported by NSF and NASA grants to ASU} $^a$, M. Bose$^a$, C. Iliadis$^b$, W. R. Hix$^{c,d}$, C. E. Woodward$^e$, R. M. Wagner$^{f,g}$\\
      $^a$Arizona State University- Earth and Space Exploration, Tempe, AZ,USA; $^b$ University of North Carolina - Physics and Astronomy, Chapel Hill, NC; $^c$Oak Ridge National Laboratory, Oak Ridge, TN;   $^d$University of Tennessee, Knoxville, TN; $^e$University of Minnesota, Minneapolis, MN; $^f$Large Binocular Telescope Observatory, Tucson, AZ; $^g$Ohio State University, Department of Astronomy, Columbus, OH \\
         
E-mail: \email{starrfield@asu.edu, maitrayee.bose@asu.edu, iliadis@physics.unc.edu, raph@ornl.gov, chickw024@gmail.com, rmw@lbto.org} }

\abstract{Cataclysmic Variables (CVs) and Symbiotic Binaries are close (or not so close) binary star systems 
which contain both a white dwarf (WD) primary and a larger cooler secondary star that typically fills its Roche Lobe.  
The cooler star is losing mass through the inner Lagrangian point of the binary and a fraction of this material is accreted by the WD.  
Here we report on our hydrodynamic studies of the thermonuclear runaway (TNR) in the accreted material that ends in a Classical Nova explosion.  
We have followed the evolution of the TNRs on both carbon-oxygen (CO) and oxygen-neon (ONe) WDs. 
We report on 3 studies in this paper.  First, simulations in which we accrete only solar matter using NOVA (our 1-D, fully implicit, hydro code).  
Second, we use MESA for similar studies in which we accrete only Solar matter and compare the results.  
Third, we accrete solar matter until the TNR is ongoing and then switch the composition in the accreted layers to a 
mixed composition: either 25\% WD and 75\% solar or 50\% WD and 50\% Solar.   
We find that the amount of accreted material is inversely proportional to the initial $^{12}$C abundance (as expected). 
Thus, accreting solar matter results in a larger amount of accreted material to fuel the outburst; much larger than in earlier 
studies where a mixed composition was assumed from the  beginning of the simulation.  
Our most important result is that all these simulations eject significantly less mass than accreted and, therefore, the WD is
growing in mass toward the Chandrasekhar Limit.}

\FullConference{The Golden Age of Cataclysmic Variables and Related Objects V (GOLDEN2019)\\
		2-7 September 2019\\
		Palermo, Italy}

\begin{document}

\section{Introduction}

The two major suggestions for the stars that explode as a Supernova of Type Ia (SN Ia) are 
the single degenerate (SD) scenario and the double degenerate (DD) scenario.  In the
SD scenario, a white dwarf (WD) in a close binary system accretes material from its companion and grows to the 
Chandrasekhar Limit.   As it nears the Limit, an explosion is initiated 
in the core.   The DD scenario
assumes a merger or collision of two WDs occurs and the resulting explosion is observed as a SN Ia outburst.   
Although the SD scenario is
capable of explaining most of the observed properties of SN Ia explosions via the delayed detonation
hypothesis  \citep[and references therein]{khokhlov_1991_aa, kasen_2009_aa, woosley_kasen_11_a,
howell_2009_ab}, there are still no binary systems that have been confirmed as
progenitors.  It is also the case that the ``zoo'' of SNe Ia types is
increasing as surveys find more and more members \citep[e.g.,][]{white_2015_aa, ruiter_2020_aa}.  
Recent results \citep{cao_2015_aa,
olling_2015_aa}, {\it in the same issue of Nature},  both favor and disfavor the
SD scenario.  Therefore, there are likely multiple channels of SN Ia progenitors and
continuing study of the SD channel is warranted.

Reviews of the various proposals for SN Ia
progenitors \citep{branch_1995_aa, ruiter_2020_aa} producing a SN Ia, and the
implications of their explosions can be found in
\cite{hillebrandt_2000_aa},
\cite{leibundgut_2000_aa,leibundgut_2001_aa}, \cite{nomoto_2003_aa},
and \cite{howell_2011_aa}.  More recently, a tremendous effort has gone into
studies of their observed 
properties \citep[cf.,][]{hillebrandt_2003_aa, howell_2010_ab, maoz_2014_aa, ruiz_2014_aa}.

New evidence in favor of continuing the studies of the SD scenario comes
from the observations of SNIa 2011fe in M101.  The exploding star was likely a carbon-oxygen (CO) WD
\citep{pnugent11} with a companion that was probably on or near the main sequence
 \citep{weidongli_2011_aa,bloom_2012_aa}.  However, EVLA
\citep{chomiuk_2012_aa} and optical \citep{bloom_2012_aa} observations
have ruled out many types of cataclysmic variable (CV).  Moreover, \cite[]{schaeferpag_2012_aa} find no
star (to stringent but not impossible limits) at the ``center'' of a SN Ia remnant in the LMC while
\cite[]{edwardspag_2012_aa} find a large number of stars near the ``center'' of a second LMC SN Ia remnant. 
In addition,
HST studies of the spatial region from which SN 2011fe exploded, suggest that
the progenitor had a luminosity less than 
$\sim10^{34}$ erg s$^{-1}$ 
\citep{graur_2014_aa}, and \cite{lundqvist_2015_aa} find no evidence for
a remnant companion in late time observations of SN 2011fe and SN 2014J.  While
these observations rule out typical Supersoft X-ray sources \citep{kahabka_1997_aa},
recent studies suggest that a CV progenitor could be fainter than that value
\citep{newsham_2014_aa, starrfield_2012_basi, starrfield_2014_aa}.

Further support for the SD channel, arises from the observations of V445
Pup (Nova Puppis 2000).  There were no signs of hydrogen in the
spectrum at any time during the outburst, especially just after discovery, but there were strong lines
of carbon, helium, and other elements in the optically thick
spectra \citep[]{wagner_2001_aa, wagner_2001_ab, henden_2001_aa,
lyke_2001_aa, woudt_2005_aa,
 woudt_2009_aa}.  Unfortunately, no one has yet done an abundance
 analysis of the early spectra  to determine an upper limit to the amount of hydrogen
 that could be present in the ejected gases.  Nevertheless,
 it is probably extremely small.  Because the system was extremely luminous before the
outburst, the secondary is thought to be a hydrogen deficient carbon
star \citep[]{woudt_2009_aa}.  Since one of the defining
characteristics of a SN Ia explosion is the absence of hydrogen or
helium in the spectrum at any time during the outburst or decline, the
existence of V445 Pup implies that mass transferring binaries exist in
which hydrogen is absent at the time of the explosion and most of the
helium is converted to carbon during the nova phase of evolution.  The latest
spectra show that this system is still in outburst and, therefore, it
has not been possible to study the underlying system \citep{tomov_2015_aa}. 

In the next section (Section \ref{codes}) we briefly discuss the two stellar evolution codes used in this study.  We follow that with
a discussion of the results with NOVA (Section \ref{novasim}) and MESA (Section \ref{mesasim}).  We continue with a discussion
of the implications of our results for SN Ia progenitors (Section \ref{progenitors}),  discuss studies with a mixed composition (Section \ref{mixed}), and end with Conclusions (Section \ref{conclusions}).
  
\section{The Stellar Evolution Codes: NOVA and MESA}
\label{codes}

This section describes the two codes used in the studies presented in this paper.  
They are NOVA \citep[][and references therein]{starrfieldpep09, starrfield_2016_aa, starrfield_2019_aa} and 
MESA \citep[][and references therein]{paxton_2011_aa, paxton_2013_aa, paxton_2015_aa, paxton_2018_aa, paxton_2019_aa}.  
NOVA is a  one-dimensional (1-D), implicit, hydrodynamic, 
computer code which includes a nuclear 
reaction network that has been extended to 187 nuclei (up to $^{64}$Ge and including the $pep$ reaction), 
the OPAL opacities
 \citep[][and references therein]{iglesias_1996_aa},
the Starlib nuclear reaction rates \citep{sallaska_2013_aa},
the Timmes equations of state \citep{timmes_1999_aa, timmes_2000_ab}, and the nuclear reaction network solver developed by \cite{hix_1999_aa}.   NOVA also includes the \cite{arnett_2010_aa} algorithm for mixing-length convection
and the Potekhin electron conduction opacities
described in \cite[][and references therein]{cassisi_2007_aa}.  
These improvements have had the effect of changing the initial structures of the WDs so that they have smaller radii and larger
surface gravities.

MESA  solves the 1D fully coupled structure and composition equations
governing stellar evolution. It is based on an implicit finite
difference scheme with adaptive mesh refinement and sophisticated
time step controls. State-of-the-art modules provide the equation of state,
opacity, nuclear reaction rates, element diffusion, boundary
conditions, and changes to the mass of the star
\citep{paxton_2011_aa, paxton_2013_aa, paxton_2015_aa, paxton_2018_aa, paxton_2019_aa}.  
MESA employs
contemporary numerical approaches, supports shared memory parallelism
based on OpenMP, and is written with present and future multi-core and
multi-thread architectures in mind.  MESA combines the robust,
efficient, thread-safe numerical and physics modules for simulations
of a wide range of stellar evolution scenarios ranging from very-low
mass to massive stars.  
The equation of state is the 2005
update of the OPAL EOS \citep{rogers_2002_aa} with an extension to lower
temperatures and densities of the SCVH EOS \citep{saumon_1995_aa}.  This
EOS is supplemented with the HELM EOS \citep{timmes_2000_ab} and the
PC EOS \citep{potekhin_2010_aa} for the regimes where they are valid. 
The choice of opacity is the OPAL opacities \citep{iglesias_1996_aa}
with the low temperature opacities of \cite{ferguson_2005_aa} and the electron 
conduction opacities of \cite{cassisi_2007_aa}.

\section{Simulations with NOVA}
\label{novasim}
 
In this section we report on the first of two studies that investigated the consequence of a WD accreting stellar material with a solar composition from a
secondary donor star. Many types of close binary systems with a WD primary have been
suggested as the progenitors for SN Ia explosions, thus we have modeled accretion onto
a wide range of WD masses with a wide range of accretion rates. 
NOVA was used to study the consequences of accretion onto WDs with masses of 
0.4M$_\odot$, 0.7M$_\odot$, 1.0M$_\odot$, 1.25M$_\odot$, and 1.35M$_\odot$ assuming only material with a solar composition.  
Two initial WD luminosities  ($4 \times 10^{-3}$ L$_\odot$ and $10^{-2}$L$_\odot$) and seven mass accretion 
rates ranging from $2 \times 10^{-11}$M$_\odot$ yr$^{-1}$ to $2 \times 10^{-6}$M$_\odot$ yr$^{-1}$ were used.  

These fully implicit, time-dependent, calculations show that the sequences exhibit the \cite{schwarzschild_1965_aa} thin shell instability.
All simulations resulted in a TNR which, in only a few cases, ejected some material, while the WD radius grew to $\sim 10^{12}$cm.   
In general, the low mass WDs did not eject any material while the high mass WDs ejected 
a small fraction of their accreted material (a maximum of $\sim4\%$ for the 1.25M$_\odot$  sequences and ranging down to $\sim0.1\%$ for the 0.7M$_\odot$ sequences).  
Therefore, the WDs are growing in mass as a result of the
accretion of solar composition material and no enrichment from core material.

\begin{figure}[htb!]
\centering
\includegraphics[width=0.9\textwidth]{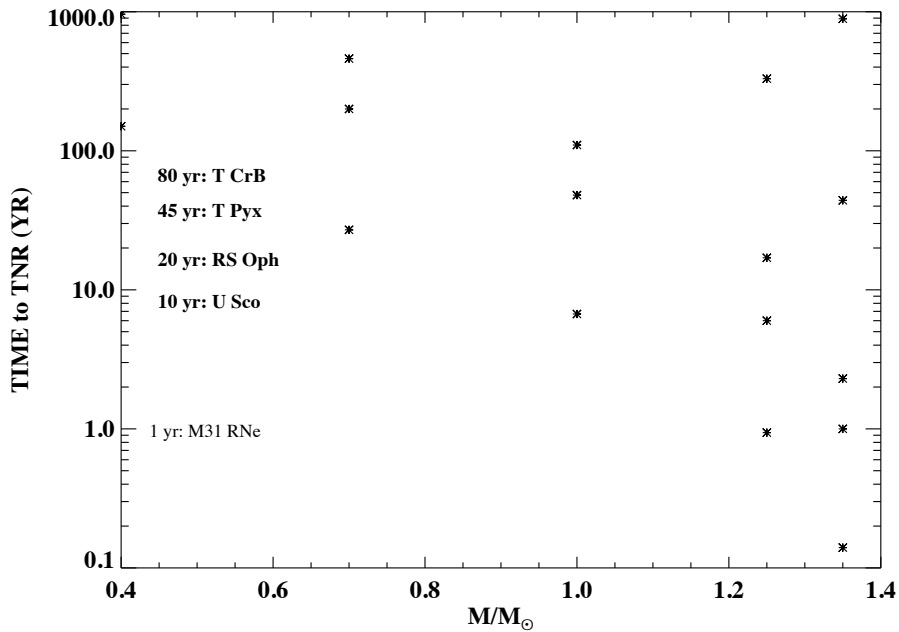}
\caption {The log of the accretion time to the TNR as a function of WD mass.  Each of the data
points is for a different \.M and the value of \.M increases downward for each WD mass.   The accretion time,
 for a given \.M decreases with WD mass because it takes less mass to initiate the TNR as the WD mass
 increases. We have identified the recurrence times of a number of recurrent novae on this plot to show that
the short times between recurrent nova outbursts implies a massive WD and a high \.M. }
\label{fig:accretiontime}
\end{figure}

Figure \ref{fig:accretiontime} shows the accretion time to the TNR for the sequences with accretion times
short enough to simulate the inter-outburst times of recurrent novae.  Note that \.M increases 
downward.  As is well known, as the WD mass increases, the accretion time decreases for the same \.M. This is because
higher mass WDs initiate the TNR with a smaller amount of accreted mass.  Given the
existence of recurrent novae and 
Symbiotic Novae with recurrence times ranging from a few years (U Sco) 
to about 20 years (RS Oph) or longer (T Pyx, V407 Cyg, and T CrB),  Figure \ref{fig:accretiontime} shows that it is possible for
recurrent novae to occur on WDs with masses as
low as 0.7M$_\odot$.  Although it is often claimed that only the most massive 
WDs have recurrence times short enough to agree with the observations of 
recurrent novae, this is not the case and basing WD mass determinations
for recurrent novae purely on short recurrence times is incorrect.  Figure \ref{fig:accretiontime} demonstrates that it is also possible 
for a recurrent nova outburst to occur on a high mass WD for an extremely broad range of \.M.  Finally, for the most massive WDs, the recurrence period
can be less than a year which suggests that the ``rapidly recurring'' recurrent nova in M31 (M31N 2008-12a) which is outbursting about once per year and
has opened up a large cavity in the ISM surrounding the system \citep{darnley_2016_aa, darnley_2017_aa, darnley_2017_ab, darnley_2019_aa, henze_2015_aa, henze_2018_aa} can be explained by our simulations.  
This system is neither X-ray nor UV luminous between outbursts and its existence argues against those investigators disparaging the SD scenario because they expect 
bright UV surroundings after a SN Ia outburst 	.  

\begin{figure}[htb!]
\centering
\includegraphics[width=0.9\textwidth]{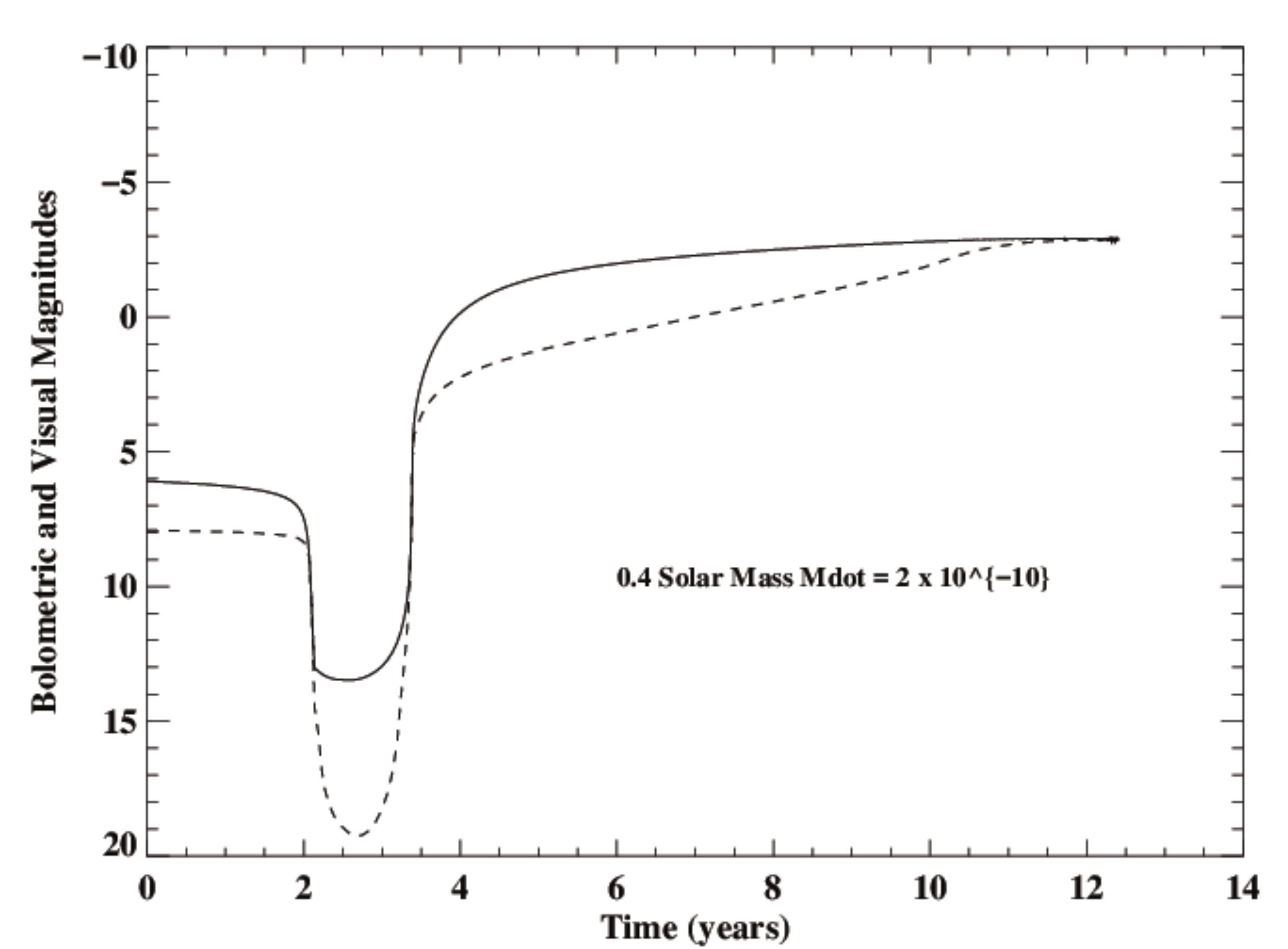}
\caption {The light curve for the 0.4 M$_\odot$  sequence. The solid line is the bolometric magnitude and the
dashed line is the visual magnitude.  The WD mass and \.M are given on the plot.  
Note that it takes more than 12 years to reach maximum light
in the visual.  The deep drop in magnitude is caused by the slowly outward moving layers cooling and fading as they
climb out of the potential well of the WD. Once the expanding layers reach a few times $10^{11}$ cm, the continuing energy produced in the nuclear burning regime heats the outer layers with a concomitant rise in luminosity.  However, they are continuing to expand and again cool allowing the visual magnitude to climb to close
to the bolometric magnitude (the bolometric correction is declining as the effective temperature of the material is declining).  The calculation is stopped when the outer radii reach $10^{12}$cm but no material has been ejected.   Given the long evolution time of this sequence, the time axis is given in years.  Those for all the
rest of the simulations are given in days. }
\label{fig:fig0p4}
\end{figure}

\begin{figure}[htb!]
\centering
\includegraphics[width=0.9\textwidth]{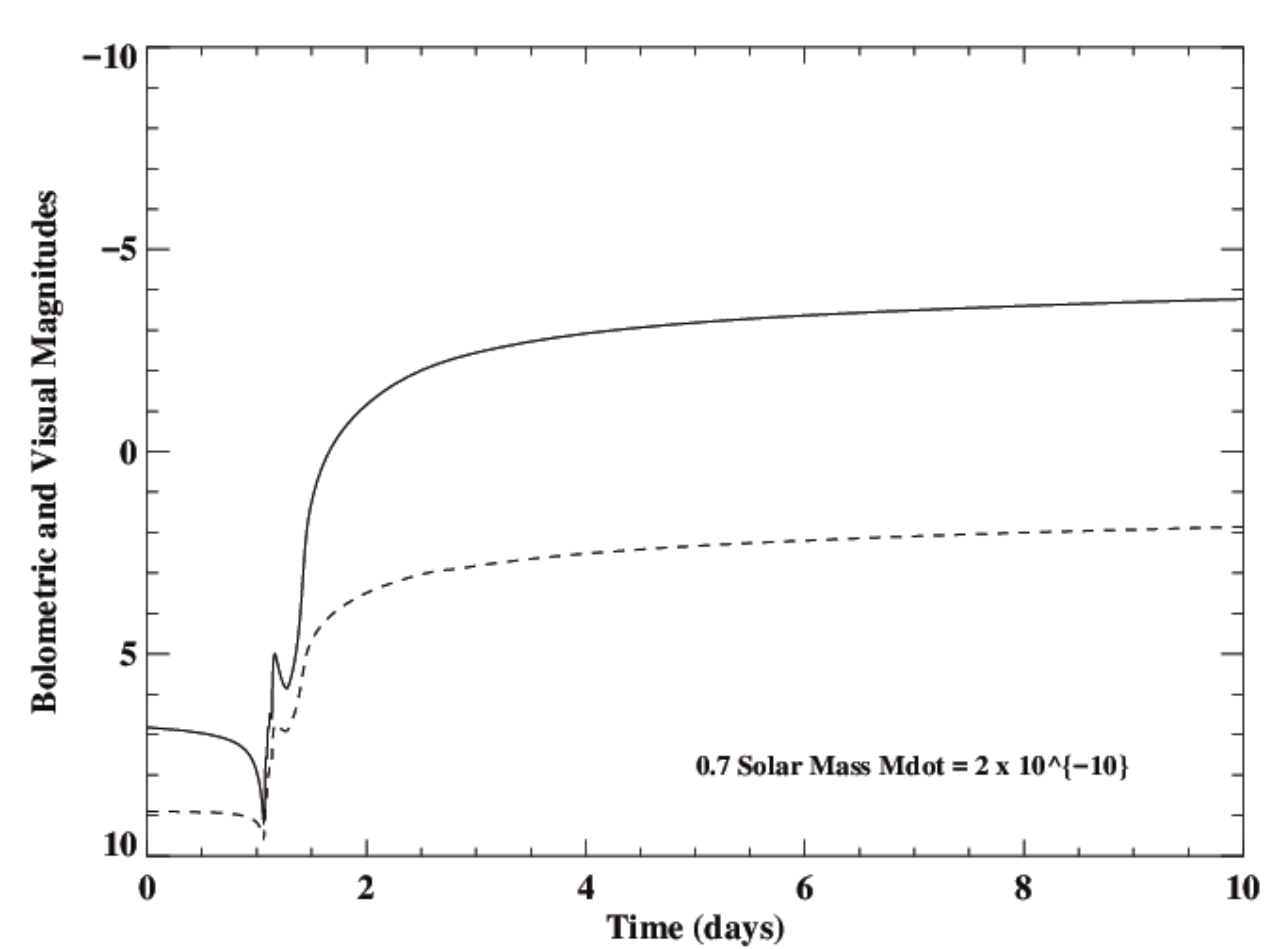}
\caption {This figure shows the light curve for a sequence with 0.7M$_\odot$.  The oscillations as the bolometric magnitude recovers
from minimum are caused by oscillations in the expansion velocity as the material climbs out of the
potential well of the WD.  The minimum magnitude is +10 which is in contrast to the minimum
magnitude of +15 realized in the 0.4 M$_\odot$ evolutionary sequence.  No material was ejected
and the calculation was stopped when the outer layers reached a radius exceeding $10^{12}$cm.  
This sequence takes less than one day for the bolometric magnitude to reach maximum. The
early rise in the light curve is emphasized to show the early structure in the rise. The
expanding material is cooling slowly and it will take days before the temperature declines to where the bolometric and visual
magnitudes are equal.}
\label{fig:fig0p7}
\end{figure}

\begin{figure}[htb!]
\centering
\includegraphics[width=0.9\textwidth]{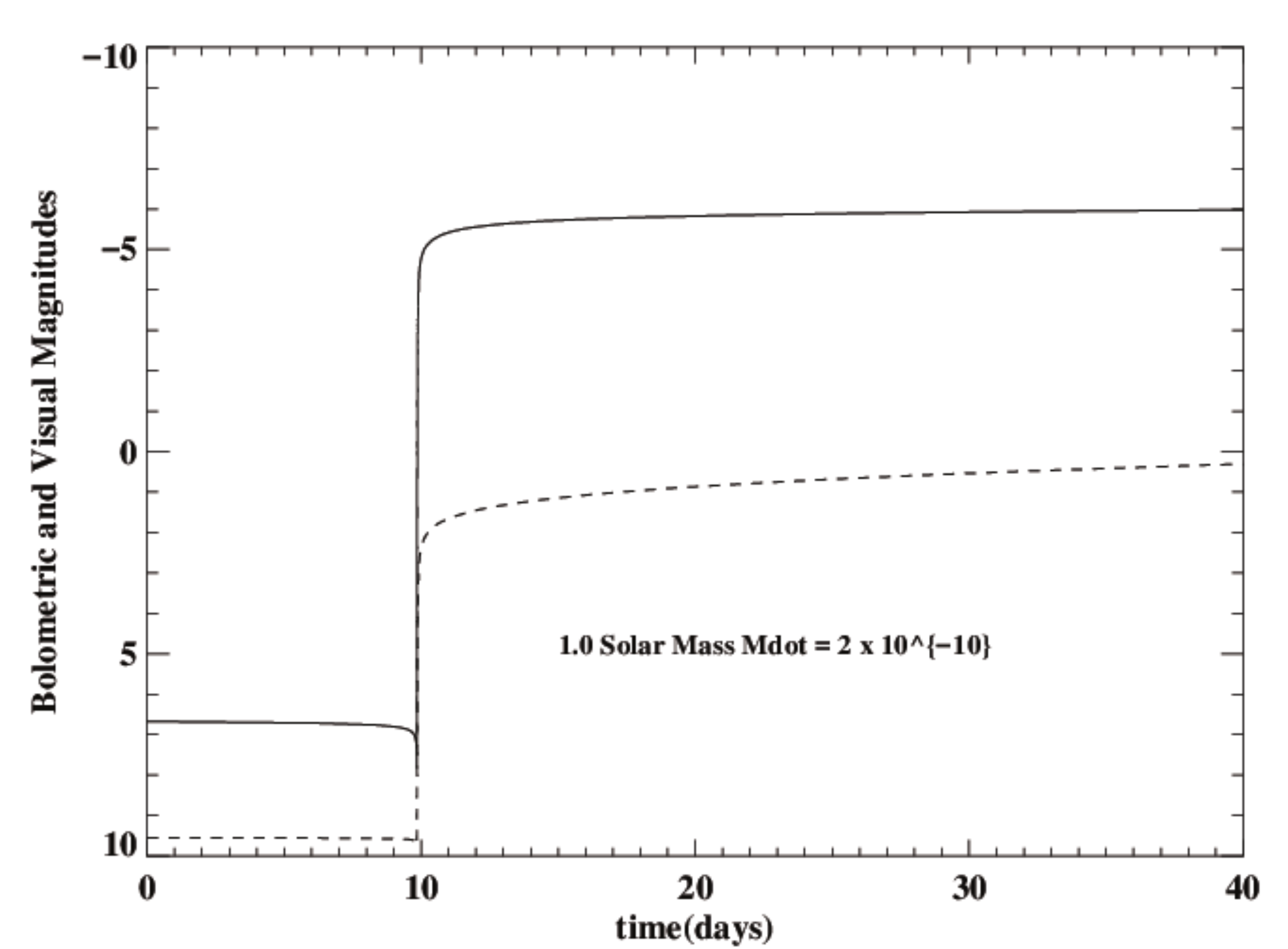}
\caption {The WD mass and \.M are given on the plot.  This figure can be compared directly with the previous
figures.  In contrast to the sequences on lower mass WDs, $8 \times 10^{-8}$ M$_\odot$ was ejected. 
This is 0.1\% of the material accreted.}
\label{fig:fig1p0}
\end{figure}

\begin{figure}[htb!]
\centering
\includegraphics[width=0.9\textwidth]{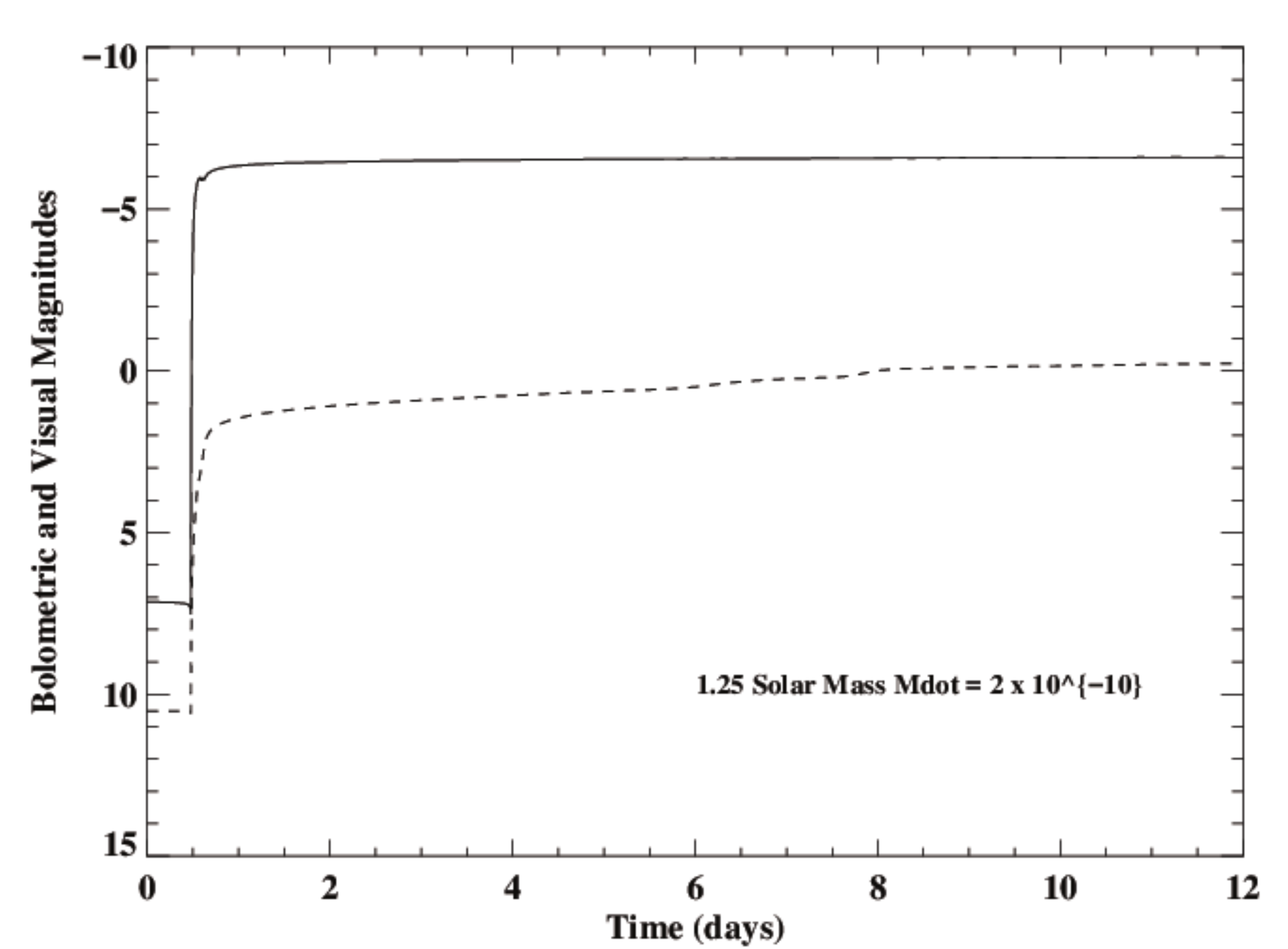}
\caption {The WD mass and \.M are given on the plot.   Note that on this WD it takes less than one day to reach 
maximum. As in the sequence at 1.0M$_\odot$ a small amount of the material was ejected (3\%).}
\label{fig:fig1p25}
\end{figure}

\begin{figure}[htb!]
\centering
\includegraphics[width=0.9\textwidth]{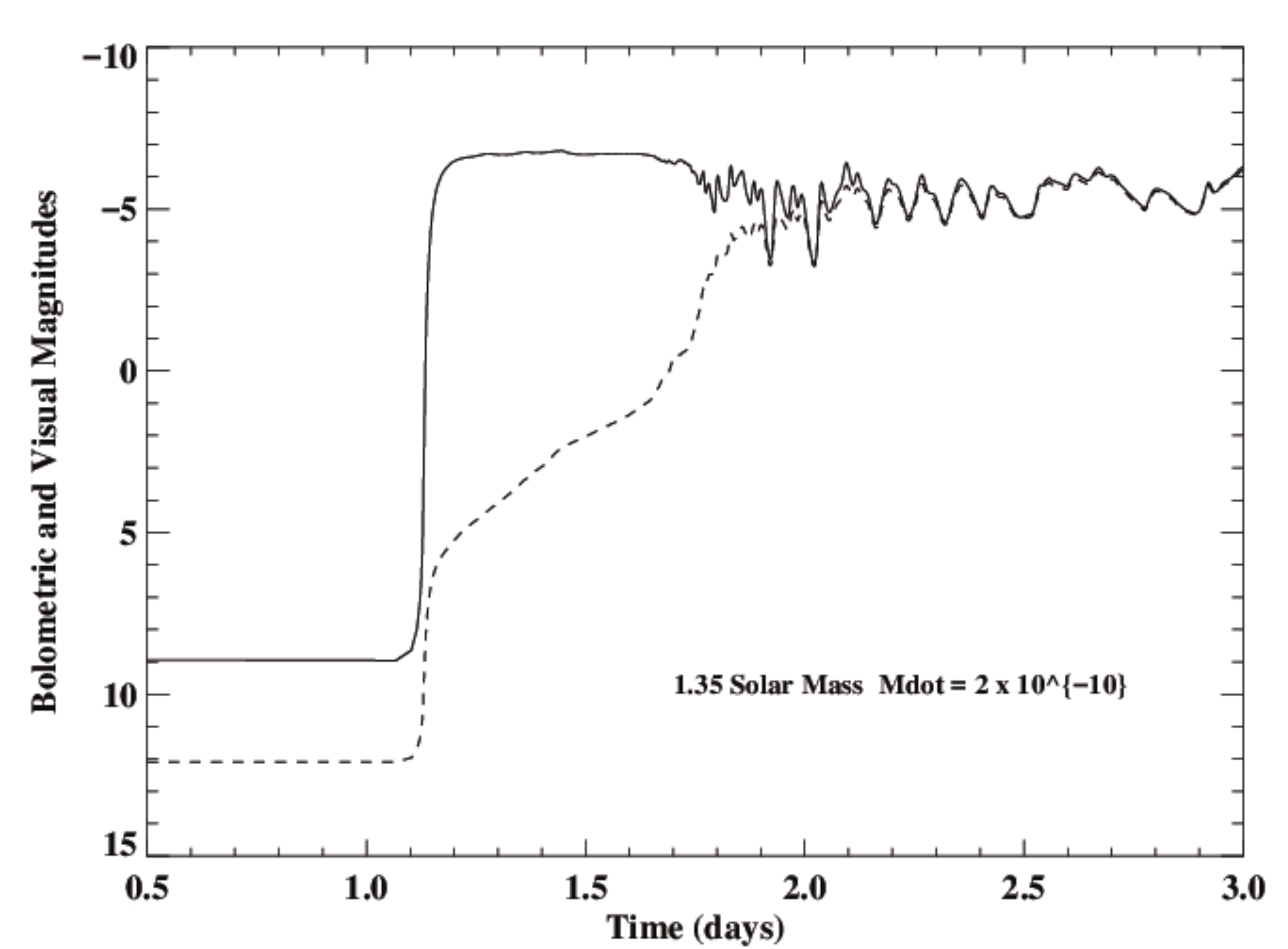}
\caption {The WD mass and \.M are given on the plot.  In contrast to the sequences on lower mass WDs, the
rise of the visual light curve is rapid and takes less than 1 day.  The oscillations occur when the material
has reached a radius of about $10^{12}$cm. No material was ejected up to the time the calculations were
stopped.}
\label{fig:fig1p35}
\end{figure}

Figures \ref{fig:fig0p4} (0.4M$_\odot$), \ref{fig:fig0p7} (0.7M$_\odot$), \ref{fig:fig1p0} (1.0M$_\odot$), \ref{fig:fig1p25} (1.25M$_\odot$), and \ref{fig:fig1p35} (1.35M$_\odot$), show the light curves for 5 simulations with the same \.M but different WD mass.  The WD mass is given on each plot.  In all cases shown, the mass accretion rate is $2 \times 10^{-10}$M$_\odot$yr$^{-1}$.  This value of \.M was chosen to obtain the largest amount of matter on the surface of the WD before the onset of the TNR which, in turn, 
should produce the highest temperatures, highest densities, highest energy generation rates, and eject the largest amount
of material.  The solid line in each plot is the evolution of the bolometric
magnitude and the V magnitude is shown as the dashed line.   While the
initial models all had luminosities $\sim 4 \times 10^{-3}$ L$_\odot$, because of their different
radii, their effective temperatures ranged from 11,500 K for the 0.4M$_\odot$ WD to 32,000 K
for the 1.35M$_\odot$ WD which can be seen in the initial differences between the values of the bolometric and
visual magnitudes.   

We stop the evolution when
the outer radius reaches $\sim 10^{12}$cm and the expanding material has become optically thin.  
The lower mass WD takes years to evolve to $\sim 10^{12}$ cm  while the higher mass WDs take only days. 
The initial long decline and slow recovery in Figure \ref{fig:fig0p4}
is caused by the conversion of some of the
internal energy, produced by ongoing nuclear burning near the surface, into the 
potential energy necessary for the material to climb out of the 
gravitational well of the WD.  The most extreme result is for the 0.4M$_\odot$ WD for which
it takes more than one year for the expanding material to recover and begin to become more 
luminous and hotter.  As the mass of the WD increases, the time
scale to the TNR decreases which can be seen in the accreted mass 
necessary to reach the TNR which ranges from $6 \times 10^{-4}$M$_\odot$
(0.4M$_\odot$) to $6 \times 10^{-6}$M$_\odot$ (1.35M$_\odot$). 

While the simulations with NOVA were done with fully 1-D hydrodynamics, NOVA is only able to
follow the first outburst on the WD.  In all cases a TNR occurred and caused the outer layers to
begin expanding to large radii.  In a few cases the outermost mass zones reached escape velocity and became
optically thin.  The amount of material lost, if any, was tabulated but the escaping zones were not removed from the
mass zoning because, even if they were escaping, they still exerted a numerical pressure 
on the underlying layers.  If this ejected material were removed, excessive mass loss would have occurred.  
In summary, the NOVA results imply
strongly that the WD is growing in mass as a result of accretion.  

Nevertheless, it is important to do simulations that allow repeated outbursts on the WD and follow the secular
evolution to see if the WD continues to grow in mass.   We do this in the next section with MESA 
because it is capable of following multiple outbursts on an accreting WD and, thereby, determining
if the WD is gaining in mass. 

\section{Simulations with MESA}
\label{mesasim}

The MESA \citep{paxton_2011_aa, paxton_2013_aa, paxton_2015_aa, paxton_2018_aa, paxton_2019_aa} studies 
followed the long term evolution of the WD as it accreted stellar material with a solar 
composition and experienced a large number of TNRs resulting in ejection, mass loss, and then the
renewal of accretion.
\textit{Therefore, the treatment of mass loss is important in following the long term behavior of the WDs, 
determining if mass growth occurs and, if so, how rapidly.} 
In order to determine the effects of different mass loss prescriptions, 
the evolution of  a 1.35M$_\odot$ WD accreting at a rate of
$1.6 \times 10^{-9}$M$_\odot$yr$^{-1}$ was followed using three different mass loss prescriptions.
These were 
(1) Eddington wind mass loss as described in \cite{shavivnj_2002_aa};
(2) determining the amount of mass
ejected when the outer layers exceed the escape velocity, are optically thin, and the surface 
radius exceeds $10^{12}$cm (this is the prescription in NOVA); and 
(3) Roche Lobe \index{Roche Lobe} overflow with an
assumed ejection rate of $10^{-6}$M$_\odot$yr$^{-1}$  at a radius of 1.0 R$_\odot$.  
Two different atmospheres were used for the Eddington
wind method. One was a grey atmosphere and the other was a WD atmosphere; both with an optical 
depth of twenty-five.  A grey atmosphere was used for the other mass loss prescriptions. 

The major differences in the results using these 3 mechanisms was that the accretion efficiency 
(defined as the mass accreted minus the mass lost divided by the mass accreted over a flash cycle) was
$\sim$13\% for the Eddington wind prescription (with little effect of the chosen atmosphere), 
$\sim$15\% to 20\% for Roche lobe overflow, and $\sim$90\% for the method used in NOVA.
Therefore, in comparison with the NOVA studies,    
the Eddington wind method ejects
a larger amount of the accreted mass than NOVA so that the WD grows 
in mass more slowly.  The Eddington wind method was used in the MESA calculations because it
is more easily implemented for following repeated TNRs.

Initial WD masses of 0.7 M$_\odot$, 1.0 M$_\odot$ and 1.35 M$_\odot$ were used.
All WDs consisted of bare CO cores (C = 0.357, O = 0.619) at the beginning of accretion.
The mass accretion rates were chosen to be
$1.6 \times 10^{-10}$ M$_\odot$yr$^{-1}$, $1.6 \times 10^{-9}$ M$_\odot$yr$^{-1}$, 
$1.6 \times 10^{-8}$ M$_\odot$yr$^{-1}$, and $1.6 \times 10^{-7}$ M$_\odot$yr$^{-1}$.
Other accretion rates were used, in order to separate different regimes of behavior, when needed. 
The composition of the accreted material was solar.  
All simulations were run for either many TNR cycles or the time required for 
long-term behavior to become evident.

\begin{figure} [htb!]
\begin{center}
\includegraphics[width=0.9\textwidth]{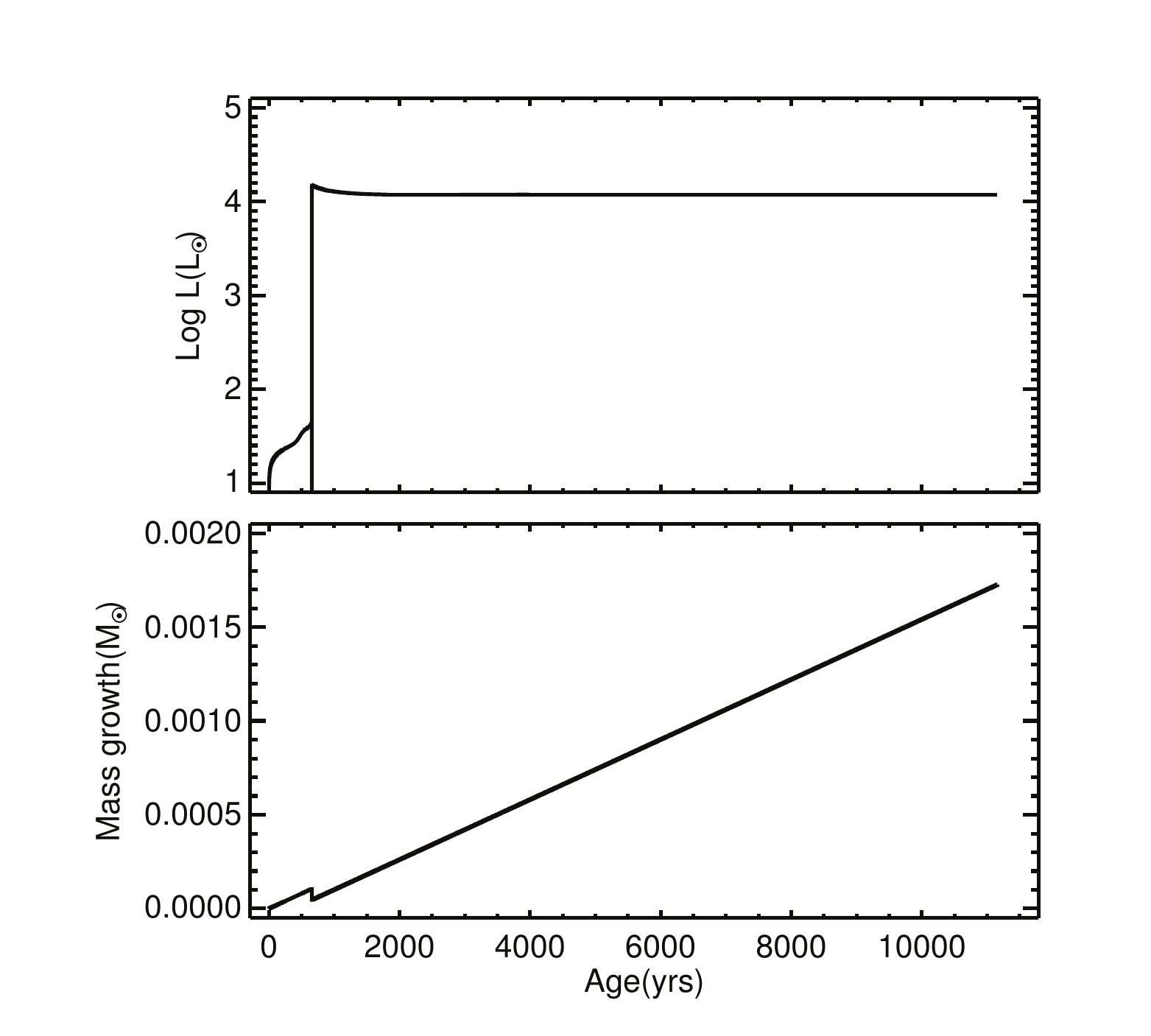}
\caption{The top panel shows the log of the luminosity as a function of time for a 0.7M$_\odot$ WD 
accreting at  $1.6 \times 10^{-7}$M$_\odot$yr$^{-1}$.  The bottom panel shows the growth in mass
as a function of time.  At this \.M there is a single TNR and then the WD slowly grows in mass.  The surface luminosity is close to that observed for the Super Soft X-ray Binary sources.
The evolution of the WD was followed for more than $10^4$ yr.}
\label{fig:fig2}
\end{center}
\end{figure}

Figure \ref{fig:fig2}, shows a  0.7 M$_\odot$ evolutionary sequence accreting 
at $1.6 \times 10^{-7}$ M$_\odot$yr$^{-1}$ that exhibits
an initial flash and then settles into an extended period of  slow growth where the
luminosity is constant.  The top panel shows the log of the luminosity as a function of time while the bottom panel
shows the growth in mass over the same time period.  
The surface luminosity is close to that observed for the
Super Soft Binary X-ray sources \citep[][and references therein]{kahabka_1997_aa} although the metallicity assumed to obtain the opacities
was solar and not that of the LMC.  
The evolution of the WD, and the slow growth in mass,  was followed for more than $10^4$ yr at which time the simulation was stopped.  
During this phase, the mass of the WD grew at a rate of $1.55 \times 10^{-7}$M$_\odot$yr$^{-1}$.  
 
\begin{figure} [htb!]
\begin{center}
\includegraphics[width=0.9\textwidth]{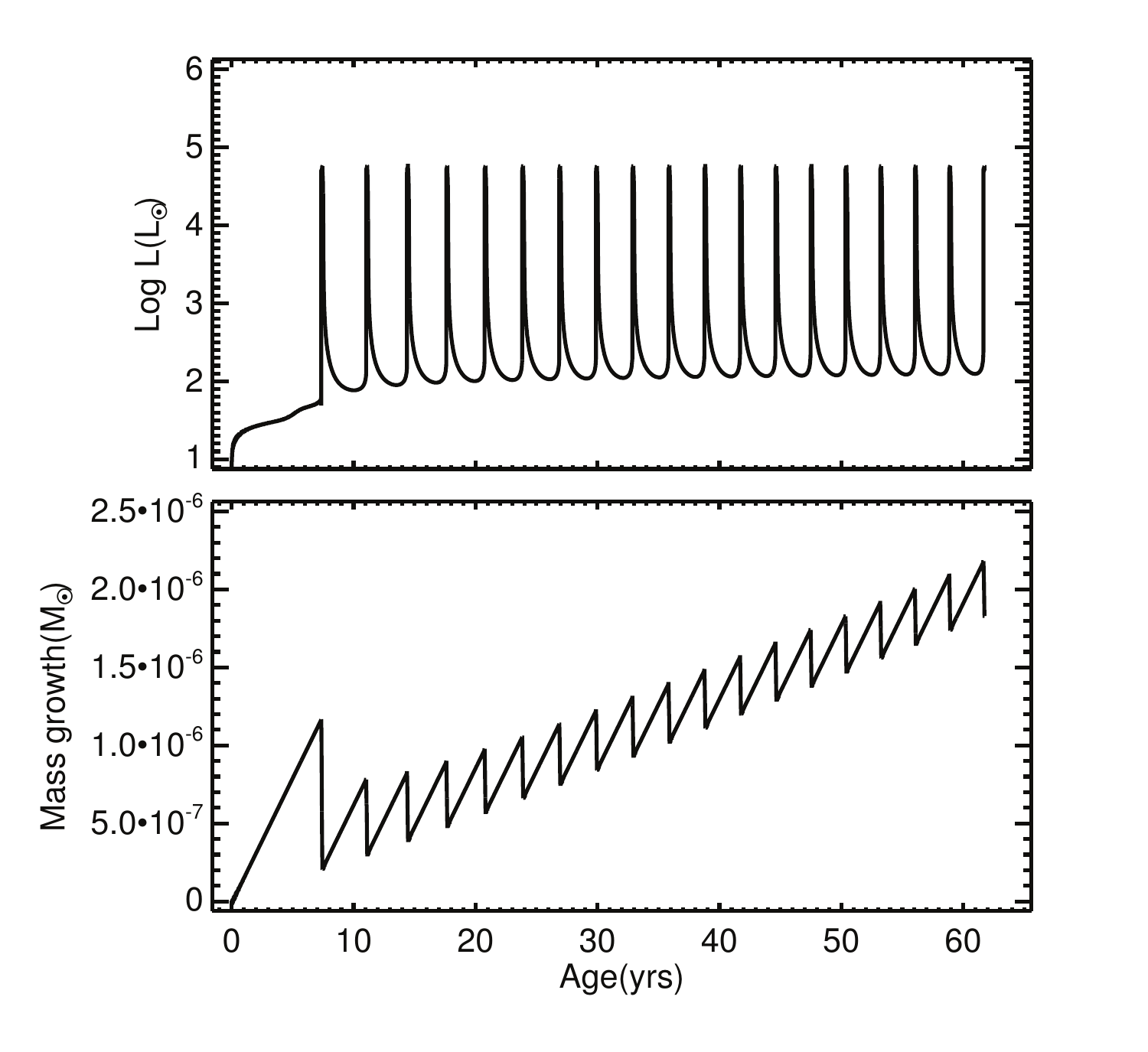}
\caption{The top panel shows the log of the luminosity as a function of time for a 1.35M$_\odot$ 
evolutionary sequence accreting at an \.M of  $1.6 \times 10^{-7}$M$_\odot$yr$^{-1}$.  The bottom panel
shows the growth in mass as a function of time.  After the initial flash, which is the strongest, the WD
slowly grows in mass.  The decrease in WD mass during each flash is caused by the mass lost when
the outermost layers exceed the Eddington Luminosity.}
\label{fig:fig1}
\end{center}
\end{figure}

\begin{figure} [htb!]
\center
\includegraphics[width=0.9\textwidth]{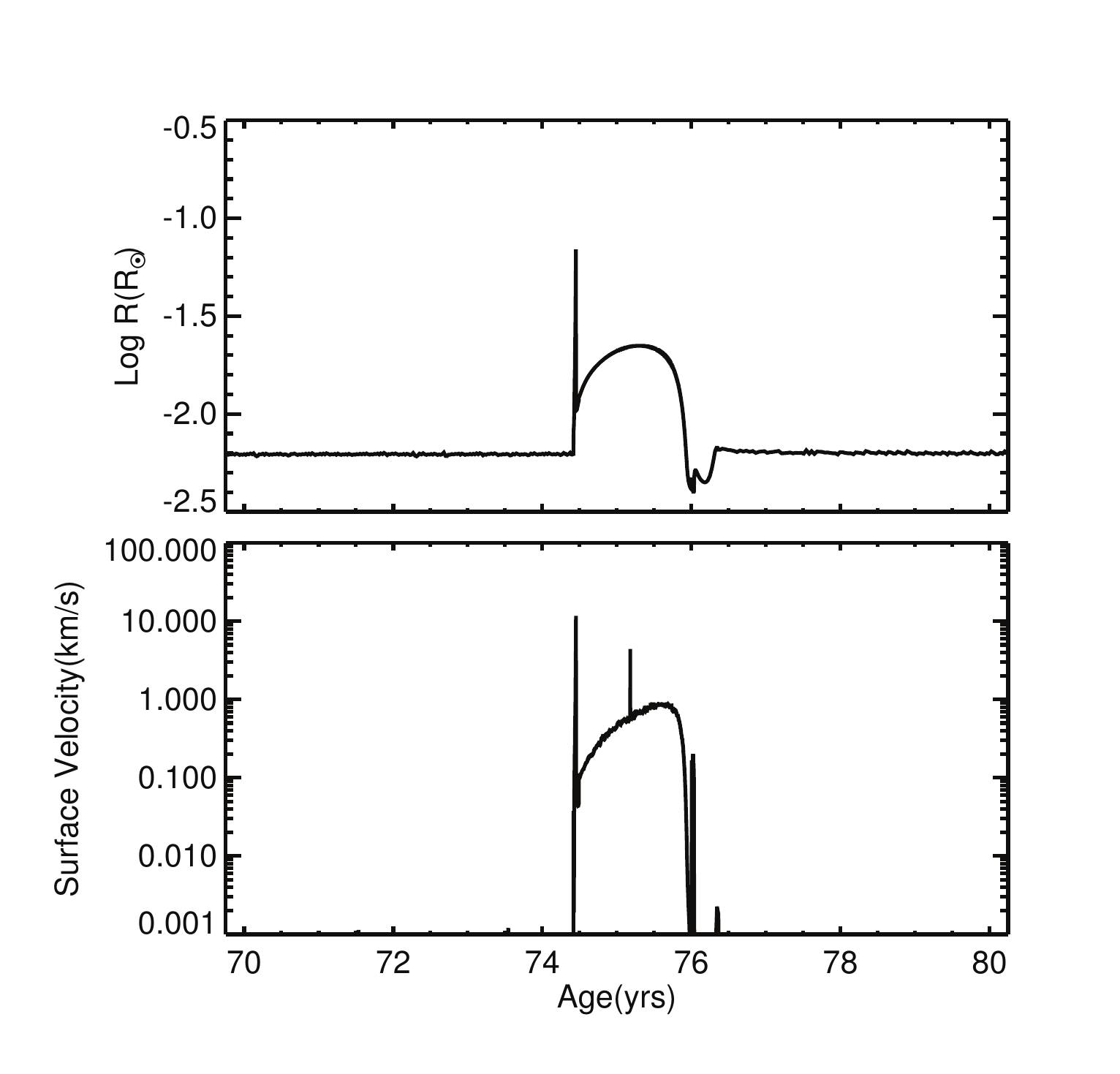}
\caption{These two panels show a helium flash that occurred on a 1.35M$_\odot$ WD accreting at a rate of
$6.4 \times 10^{-7}$M$_\odot$yr$^{-1}$.  The upper panel shows the log of the radius (in units of the
Solar radius) as a function of time and the lower panel shows the velocity of the surface layers (km s$^{-1}$) 
over the same time period.  The surface material is never able to climb very far out of the potential well of
the WD and never reaches very large velocities. }
\label{fig:fig6}
\end{figure}

Figure \ref{fig:fig1} shows both the surface luminosity versus time (top
panel) and the mass growth (bottom panel) for a 
1.35 M$_\odot$ WD accreting at $1.6 \times 10^{-7}$M$_\odot$yr$^{-1}$.  
After the initial growth to the first TNR, 
the model settles into a recurring pattern of  mass accretion
resulting in TNRs  which eject a fraction of the accreted mass via an Eddington
wind.  The positive slope shows that the WD is growing in mass at a rate of 
$\sim 3.0 \times 10^{-8}$M$_\odot$yr$^{-1}$ for an accretion rate of 
$1.6 \times 10^{-7}$M$_\odot$yr$^{-1}$.  The recurrence time for this sequence is almost short 
enough to agree with that of the M31 recurrent nova that is outbursting almost once per
year \citep{darnley_2016_aa, darnley_2017_aa, darnley_2017_ab, darnley_2019_aa, henze_2015_aa, henze_2018_aa}.

While the evolution shown in Figure \ref{fig:fig1} implies that a 1.35M$_\odot$ simulation can grow
in mass with repeated hydrogen flashes, it is also interesting to see if accretion at a higher
rate will cause the WD to grow in radius to that of a red giant as proposed by
\citep{fujimoto_1982_ab, fujimoto_1982_aa}. 
Therefore, a higher \.M was chosen ($6.4 \times 10^{-7}$M$_\odot$yr$^{-1}$) and 
while helium flashes \index{helium flash}occurred they did not eject any material. 
This is shown in Figure \ref{fig:fig6}.   
After approximately seventy-five years of accretion,  the WD undergoes
a helium flash. 
The radius of the WD peaks at 0.056 R$_\odot$ but the peak surface velocity reaches only
10 km s$^{-1}$.

If this WD were in a CV system with a secondary 
that had a mass of 0.7M$_\odot$ and a radius of 0.75 R$_\odot$, 
it would fill its Roche Lobe if the system had a semi-major axis of 
2.32R$_\odot$ and a period of 6.9 hr.  This gives a Roche Lobe radius for the 
WD primary of 1.01R$_\odot$.  However, the radius in this simulation never
exceeded more than about 6\% of this value during the helium flash.  Therefore,
such helium flashes do not stop the growth in the mass of the WD from 
either Roche Lobe overflow or envelope ejection.  After the initial helium flash, the WD undergoes
succeeding helium flashes roughly every 75 years.  The WD continues to grow in mass
at a rate of $2.6 \times 10^{-7}$ M$_\odot$yr$^{-1}$.
The helium flashes are not violent because mass is
being lost via an Eddington wind.  Thus, a helium flash is not the dynamic mass ejection event
that occurs with the mass loss prescription used in NOVA.   In complementary helium accretion simulations with NOVA,
large radii resulted from the initial TNR and further evolution was halted.

\begin{figure} [htb!]
\center
\includegraphics[width=0.9\textwidth]{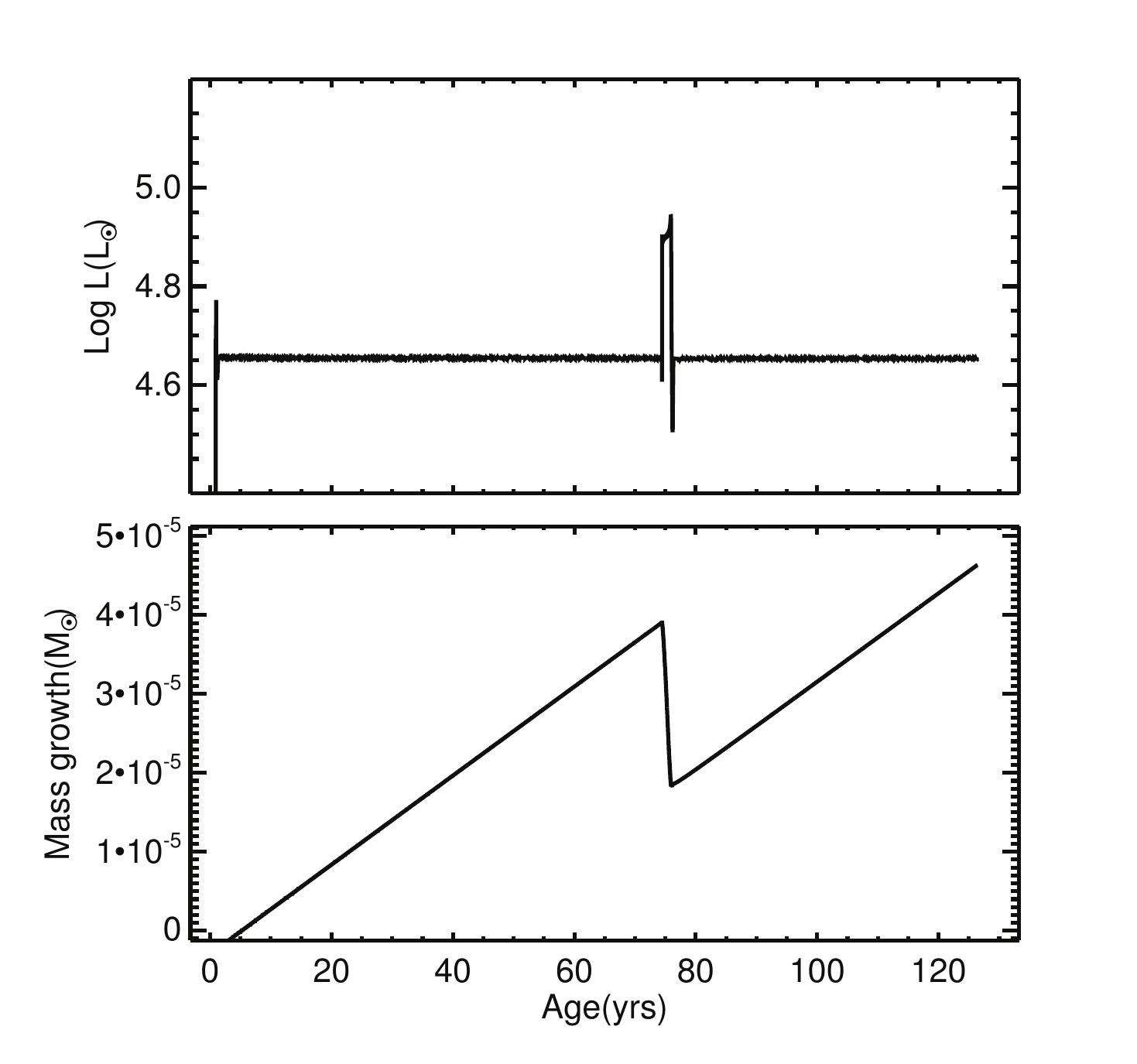}
\caption{Same as for Figure \ref{fig:fig2} but for a 1.0M$_\odot$ WD accreting at  $6.4 \times 10^{-7}$M$_\odot$yr$^{-1}$.
The WD exhibits steady growth interrupted by a helium flash which does not eject all the accreted material so that the WD is  growing in mass. }
\label{fig:fig3}
\end{figure}

\begin{figure} [htb!]
\center
\includegraphics[width=0.9\textwidth]{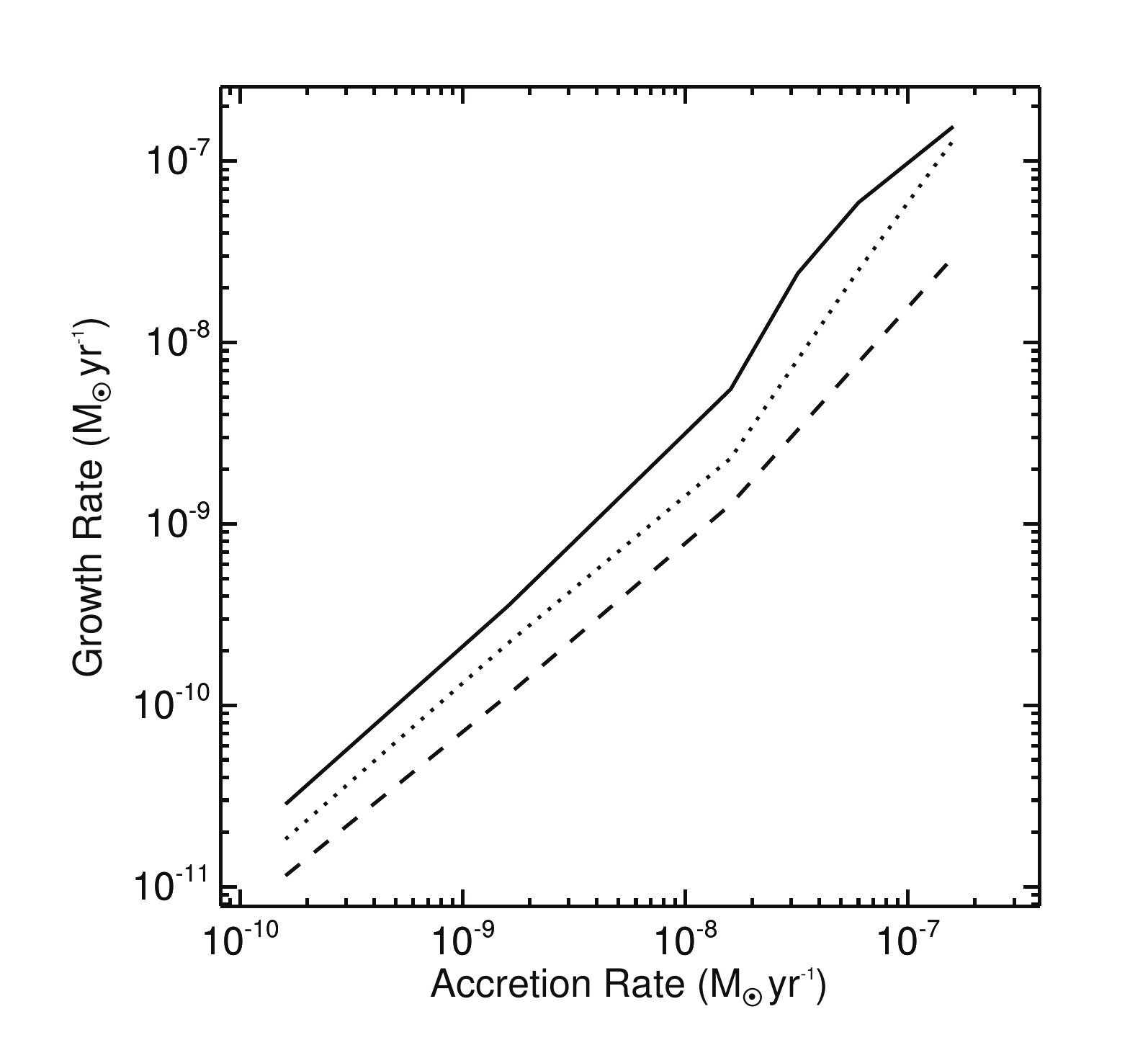}
\caption{The growth rate in  M$_\odot$yr$^{-1}$ as a function of \.M for 0.7M$_\odot$ (solid line), 1.0M$_\odot$
(dotted line), and 1.35M$_\odot$ (dashed line) WDs.  The lowest mass WDs are growing faster than the most massive WDs for the same \.M 
since they accrete more mass and eject no mass as a consequence of the TNR. }
\label{fig:fig4a}
\end{figure}

In Figure \ref{fig:fig3} the evolution of a 1.0 M$_\odot$ WD also accreting at  $6.4 \times 10^{-7}$M$_\odot$yr$^{-1}$ is shown.
The WD exhibits steady growth interrupted by a helium flash that again does not eject all the accreted material so that the WD is growing in mass.  
At lower accretion rates, the 0.7 M$_\odot$, 1.0 M$_\odot$, and 1.35M$_\odot$ sequences also go through TNRs and
mass loss events and the WD grows in mass.   This is shown in Figure \ref{fig:fig4a} which gives the
growth rate in  M$_\odot$yr$^{-1}$ as a function of \.M for 0.7M$_\odot$ (solid line), 1.0M$_\odot$
(dotted line), and 1.35M$_\odot$ (dashed line) WDs.  The lowest mass WDs are growing faster than the most massive WDs for the same \.M. 

The growth rates (Figure \ref{fig:fig4a}) are much larger for the higher accretion rates at a given WD mass. 
For a given accretion rate, the lower
mass WDs grow in mass at a greater rate than the more massive
WDs.  This is because the lower mass WDs must accrete more mass to reach the initial conditions for a
TNR and, since they produce less nuclear energy, less mass is ejected.  
The 0.70M$_\odot$ and 1.0M$_\odot$ WDs, for accretion
rates greater than $\sim 10^{-7}$M$_\odot$yr$^{-1}$, eventually become 
red giants at which point the simulation is ended.  However, the 1.35M$_\odot$
sequences, although exhibiting helium flashes, continue their mass growth at the
higher accretion rates and never grow to red giant radii.

\begin{figure} [htb!]
\center
\includegraphics[width=0.9\textwidth]{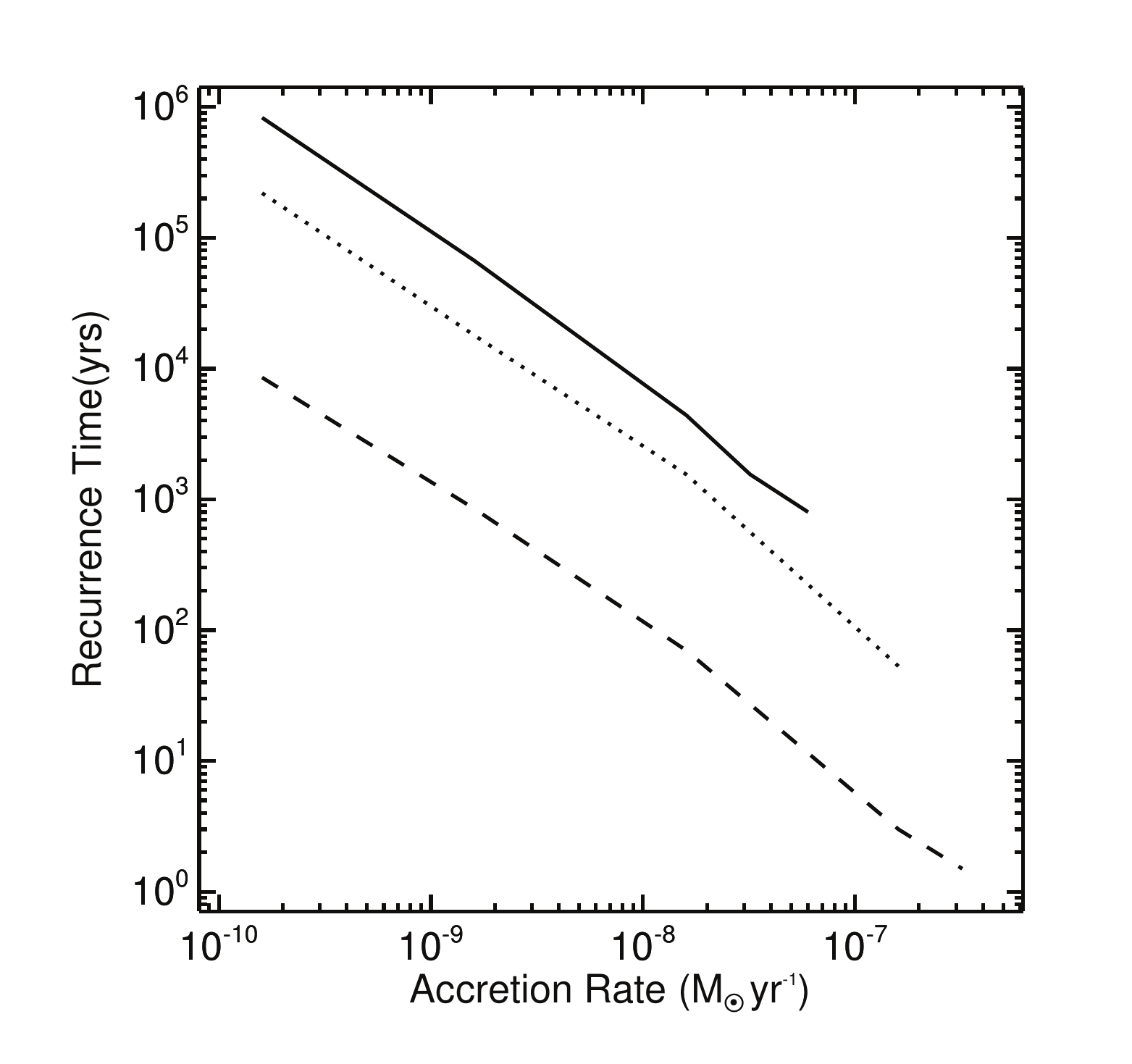}
\caption{The recurrence time in years as a function of \.M for 0.7M$_\odot$ (solid line), 1.0M$_\odot$
(dotted line), and 1.35M$_\odot$ (dashed line) WDs.  This figure is to be compared with Figure \ref{fig:accretiontime} 
which exhibits the same behavior and nearly the same recurrence times but are the results from
calculations done with NOVA and not MESA.}
\label{fig:fig5}
\end{figure}

Using the results shown in Figure \ref{fig:fig4a}, we estimate the timescale for the growth of the WDs in CV systems
to reach the Chandrasekhar Limit.  The sequences 
that grow in radius to red giants (\.M greater than $\sim 10^{-7}$M$_\odot$yr$^{-1}$ on lower mass WDs)
and also the lower accretion rate sequences which exhibit the most violent hydrogen flashes
(\.M less than $\sim 10^{-8}$M$_\odot$yr$^{-1}$) are ignored in these figures.
Using these constraints, the time for a 0.7M$_\odot$ WD to reach the Chandrasekhar Limit 
\index{Chandrasekhar Limit} ranges
from $1.5 \times 10^{7}$yr to $ 4.3 \times 10^{8}$yr.  The 1.0M$_\odot$ WD will take 
$1.2 \times 10^{7}$yr to $ 3.3 \times 10^{8}$yr, and the
1.35M$_\odot$ WD requires $5.6 \times 10^{6}$yr to $ 1.1 \times 10^{8}$yr.   
The shortest and longest evolution times vary by two orders of magnitude. 

Figure \ref{fig:fig5} shows the recurrence time of the flashes versus the accretion rate (excluding
the sequences that terminate as a red giant). For the 0.70M$_\odot$ WD,  the recurrence time
spans some three orders of magnitude with the lowest accretion rate exhibiting the longest
recurrence time.  The 1.00M$_\odot$ and 1.35M$_\odot$ WD's exhibit the same behavior with each higher
mass being approximately one order of magnitude less in recurrence time at a given accretion
rate.  The evolutionary sequences span recurrence times from $\sim 10^6$ yr down to a year or less.

\begin{figure} [htb!]
\center
\includegraphics[width=0.9\textwidth]{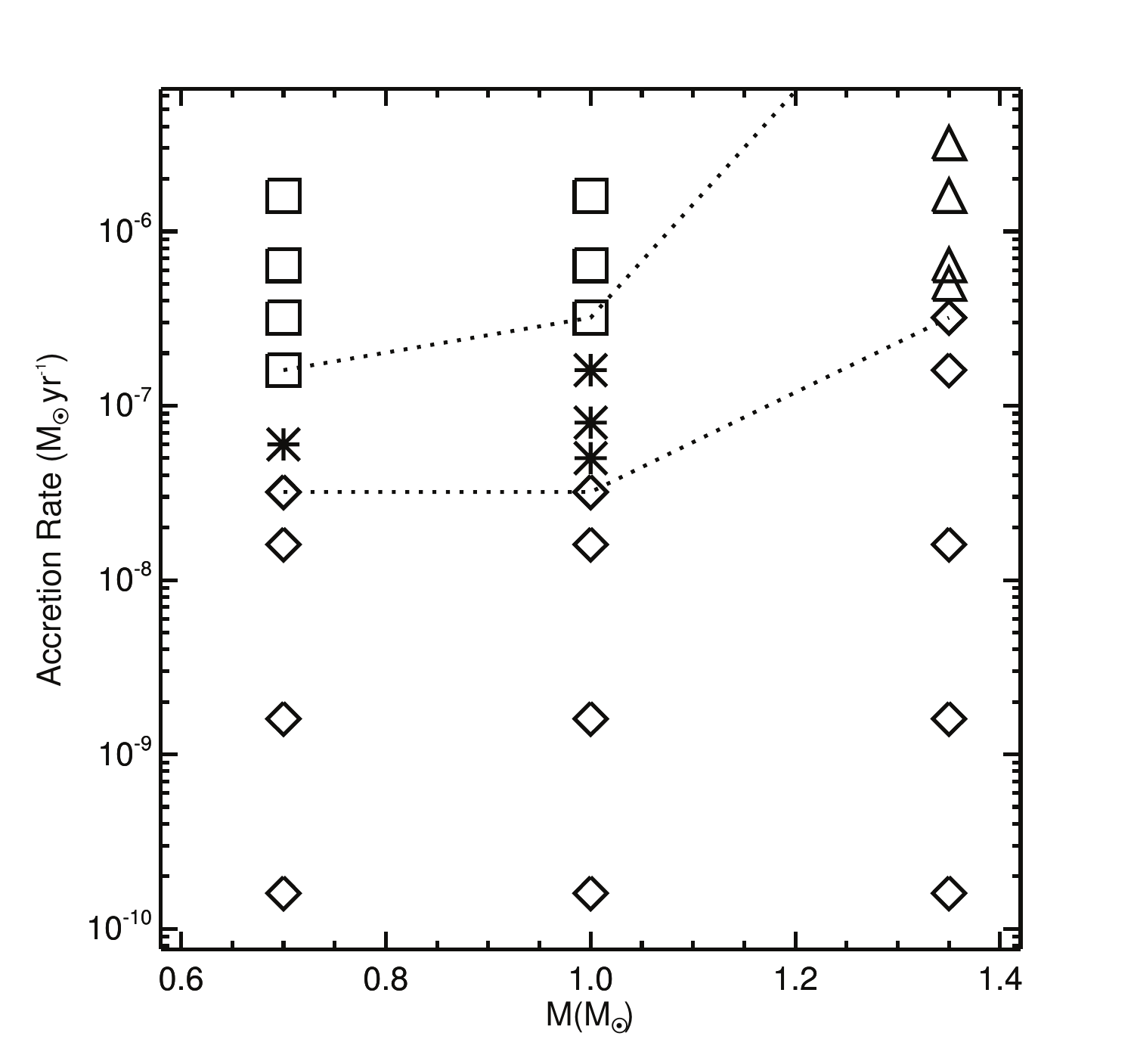}
\caption{This figure shows the
fate of accreting WDs versus their mass. The symbols indicate the sequences that evolve to red giants (squares),
those that grow steadily in mass followed by hydrogen flashes that eject only a fraction of the accreted material
(asterisk), recurrent hydrogen flashes in which the WD is growing steadily in mass (diamonds), and
slow growth in WD mass interrupted by helium flashes (triangles).  The dotted lines demarcate the 3 regions
of importance on this diagram - see the text.}
\label{fig:fig7}
\end{figure}

A summary of the MESA  results is given in Figure \ref{fig:fig7} which 
has three main regions. 
At low accretion rates, all WD masses undergo a 
TNR.  While mass is lost during the peak of the flash it is much less than that accreted to initiate
the flash.  This is the region below the lower dotted line in Figure \ref{fig:fig7} and each of the simulations 
is indicated by a ``diamond''.   At the highest accretion rates for the
0.70$_\odot$ and 1.00M$_\odot$ WDs,
the evolutionary sequences, after an initial TNR enter a regime of slow growth in mass which 
is eventually terminated by the WD growing to red giant dimensions.  The simulation is stopped 
because there is no accretion at these radii.  These evolutionary sequences exist
above the top dotted line in Figure \ref{fig:fig7}.   Nevertheless, the enlarged WD extends
past the Roche Lobe \index{Roche Lobe}radius of observed \index{Cataclysmic Variables}
CVs and common envelope evolution could eject
the outer layers and accretion begin again.  However the \.M for these sequences is larger
than typically observed for CVs.   This behavior does not occur
for 1.35M$_\odot$ WDs.  They experience helium flashes \index{helium flash} which do not eject all the accreted material
and the WD continues to grow in mass.

The third regime, intermediate between the two dotted lines, is where slow
growth in WD mass occurs for long times but the sequences eventually enter a regime of TNRs 
\index{thermonuclear runaway} characteristic of a somewhat lower accretion rate (where 
the flashes occurred at the onset of accretion).   
The intermediate behavior is approximately
bounded by accretion rates of  $5.0 \times 10^{-8}$M$_\odot$yr$^{-1}$
and $10^{-7}$M$_\odot$yr$^{-1}$.     In virtually all sequences shown in Figure \ref{fig:fig7},
the WD is growing in mass. 

Finally, it is commonly assumed that a CO WD should not have a mass exceeding $\sim 1.15 $M$_\odot$ \citep{iben_1991_aa, ritossa_1996_aa, iben_1997_aa}. 
Our simulations in this paper, however, suggest that WDs are growing in mass,
so that there should be massive CO WDs in CN systems.  An example of this class is Nova LMC 1991, a CO nova,
which exhibited a super Eddington luminosity for more than 2 weeks  \citep{schwarz_2001_aa} likely requiring  a WD mass exceeding $\simeq 1.35$M$_\odot$.  
Moreover, the WDs in four of the nearest CVs
[U Gem: 1.2 M$_\odot$ \citep[][]{Echevarria_2007_aa}, 
SS Cyg: 0.8 M$_\odot$ \citep[][]{Sion_2010_SSCyg_aa}, 
IP Peg: 1.16 M$_\odot$ \citep[][]{copperwheat_2010_aa}, and 
Z Cam: 0.99 M$_\odot$ \citep[][]{Shafter_1983_aa}] are more massive than the canonical value for single WDs of 0.6 M$_\odot$ \citep{sion_1986_aa}.   
In addition, \cite{sion_2019_aa} report a WD mass for the recurrent nova CI Aql of 0.98 M$_\odot$, \cite{shara_2018_ab} report that the mean WD mass for
82 Galactic CNe is 1.13 M$_\odot$ and for 10 recurrent novae is 1.31 M$_\odot$, while \cite{selvelli_2019_aa} analyzed
18 old CNe, using data from both IUE and Gaia, and report that many WDs in CNe have masses above the canonical value for single WDs.

\section{CNe and CVs are probably one Channel of SN Ia progenitors}
\label{progenitors}

Although of great importance to both galactic chemical evolution and, in addition, as probes of  
the evolution of the universe, the progenitors of SN Ia explosions are as yet unknown. 
Originally, the SD scenario, with the WD accreting from the secondary and growing in mass toward the Chandrasekhar Limit,
was preferred.  However, this scenario is now disfavored by many.  The other scenario, the DD
scenario, which involves either a merger or collision between two CO WDs, is now thought to be the major channel for SN Ia explosions. 
The switch in the preferred paradigm is caused by a number of perceived problems with the SD scenario that need to be understood.  
In this section we discuss four of those problems and show that they are, in fact, not problems at all.

The first major problem, which is directly relevant to the simulations presented in the earlier sections,
is the common assumption, based on the analyses of the ejecta abundances and ejecta masses of
CNe outbursts that the WD is \textit{decreasing} in mass as a consequence of the TNR and resulting explosion.  
However, that assumption is incorrect for solar accretion as we have now shown in earlier sections of this manuscript.  
In addition, \cite{starrfield_2019_aa} along with \cite{hillman_2015_ab, hillman_2015_ac, hillman_2016_aa} show that the WDs in CO CNe are also
growing in mass.  Moreover, we have now done simulations of ONe CNe and find that 
the WD in those systems is also gaining in mass (Starrfield et al. 2020, in preparation).
Therefore for CVs in general the consequence of accretion are a growth in mass of the WD.
One concern, nevertheless, is that the large number of CVs in the galaxy may
result in too many SN Ia explosions.  However, the mass of the secondary also determines the 
ultimate consequences of the evolution.
It is possible that in many CV systems the secondary has too little mass and the outbursts will stop 
and the evolution will end before the WD has reached the Chandrasekhar Limit.

The second perceived problem is due to the {\it interpretation} of the calculations of \cite{nomoto_1982_aa} and \cite{ fujimoto_1982_aa, fujimoto_1982_ab}. 
A reproduction of their results can be found as Figure 5 in \cite{kahabka_1997_aa}  which shows that the space describing the consequences of mass accretion rate as a function of the mass of the
accreting WD can be divided into three regions.  Our version of their Figure 5 is
Figure \ref{fig:fig7} and the data that we plot in Figure \ref{fig:fig7} is discussed in the last section.   
According to the results plotted in their Figure 5, for the lowest mass accretion rates, at all WD masses, they predict that accretion results in hydrogen flashes that resemble those of CNe and the WD is assumed to be losing mass.  
However,  the purpose of this paper has been to provide a broad range of simulations at various \.M and WD mass (using two independent stellar evolution codes) which show that a WD accreting at low rates is gaining in mass.  In addition, \cite{hillman_2015_ab, hillman_2015_ac, hillman_2016_aa} and we (Figure \ref{fig:fig7} ) have investigated
the consequences of accretion at higher rates and again find that the WD is growing in mass.  Thus, mass accreting systems 
with a broad range in WD mass and  \.M must be included in the classes of SN Ia progenitors.

A third problem relates to the upper region on the \cite{nomoto_1982_aa} and the \cite{fujimoto_1982_aa, fujimoto_1982_ab} plot, which shows the results for the highest accretion rates and predicts for all WD masses
that the radius of the WD will grow rapidly to red giant dimensions, accretion will be halted, and any further evolution will
await the collapse of the extended layers. These systems, therefore, cannot be SN Ia progenitors.   
However, we have done extensive studies of solar accretion onto WDs and our fully hydrodynamic studies show,
for the highest mass accretion rates on the most massive WDs, steady hydrogen burning (see below) is occurring followed by recurring helium
flashes.  The helium flashes do not result in ejection and the WDs are again growing in mass. \cite{hillman_2015_ab, hillman_2016_aa} also
report that helium flashes do not eject material.

The fourth problem is based on the existence of the intermediate regime identified by \cite{nomoto_1982_aa} and \cite{fujimoto_1982_aa, fujimoto_1982_ab}, 
where the material is predicted to burn steadily at the rate it is accreted.  The central \.M of this region is nominally 
$\sim 3 \times 10^{-7}$M$_\odot$ yr$^{-1}$ and it does have a slight variation with WD mass.   
Those systems that are accreting at the steady nuclear burning rate are supposedly evolving horizontally in this plot towards higher WD mass 
and, by some {\it unknown} mechanism, the mass transfer in the binary system is stuck in this mass accretion range.  
\cite{vandenheuvel_1992_aa} predicted that it was only the systems in this region that
were SN Ia progenitors via the SD scenario, They identified the Super Soft X-ray sources (SSS) as those systems, based on their luminosities and effective temperatures. 
The SSS  are $\it luminous$, massive, WDs  discovered by ROSAT \citep{Trumper_1991_aa}.  
They are binaries, with luminosities L$_* \sim 10^{37-38}$erg s$^{-1}$
and effective temperatures ranging from $3 - 7 \times 10^5$K \citep{cowley_1998_aa, branch_1995_aa, kahabka_1997_aa}.  

However, in more recent studies of accretion without mixing, an expanded study of the stability of thin shells can be found in \cite[][and references therein]{yoon_2004_aa}, who investigated
the accretion of helium-rich and hydrogen-rich material onto WDs.  Their results show that sequences in the steady nuclear burning regime begin in a stable region, 
but with continued accretion, evolve into instability.   In addition, their study shows that the evolutionary sequences
at these \.M exhibit the \cite{schwarzschild_1965_aa} thin shell instability, 
which implies that steady burning does not occur.  We identify these systems, therefore, with those CVs (dwarf, recurrent, symbiotic
novae) that show no core material either on the surface of the WD or in their ejecta. 

Given that the SSS were the 
only systems that were predicted to be SD Ia progenitors, it was expected that they would be detected by consequences of the long
periods of luminous X-ray and UV emission on the surrounding ISM.  In addition, this extreme emission should still be evident in the ISM surrounding recent SN
Ia explosions.   For example, \cite{graur_2019_aa} state: "For the WD to efficiently grow in mass, the accreted hydrogen must undergo stable nuclear-burning on its surface. This means the progenitor system will be a luminous source of soft X-ray emission \citep[a supersoft X-ray source, SSS,][]{vandenheuvel_1992_aa} for at least some period of time before the explosion."  Similar statements can also be found in \cite{gilfanov_2010_aa} and \cite{kuuttila_2019_aa}. 
Such emission has not been found and the absence of evidence has been used to eliminate the
SD scenario even in the most recent studies.  However, observations of CNe and CVs, which we now identify as
possible SN Ia progenitors, show that they do not spend a large amount of time at high luminosities and effective temperatures.  

Moreover, some recurrent novae are repeating sufficiently often that their WDs
must have grown in mass so that they are now close to the Chandrasekhar Limit.  One such system is the ``rapidly recurring'' 
recurrent nova in M31 (M31N 2008-12a) which is outbursting about once per year and
has opened up a large cavity in the ISM surrounding the system \citep{darnley_2016_aa, darnley_2017_aa, darnley_2017_ab, darnley_2019_aa, henze_2015_aa, henze_2018_aa}.  
It is neither X-ray nor UV luminous between outbursts.  

\section{New Studies with Mixed Compositions}
\label{mixed}

In the previous sections we described the consequences of TNRs on WDs where no mixing of the accreted solar material
with WD material was assumed.  In this section we briefly describe new studies of the consequences of TNRs that result when mixing
of accreted material with WD core material occurs.  While almost all of the earlier studies assumed that the accreting 
material mixed immediately with WD material, we have now altered how and when we assume the mixing occurs.  The
reason is that if we assume mixing occurs from the beginning of the simulation the resulting simulations do not agree
with the observations.  The basic reason is that once the accreting envelope has been enriched, the increased opacity
traps more of the heat from compression and nuclear burning in the accreted layers.  This, in turns, causes the temperature
to rise more rapidly per unit accreted material and the TNR occurs earlier with less material accreted than if no mixing had
occurred \citep{starrfield_1998_aa, starrfield_2016_aa}.

Fortunately, the recent multi-dimensional studies of convection in the accreted layers of WDs \citep[][and references therein]{casanova_2010_aa,
 casanova_2010_ab, casanova_2011_aa, casanova_2011_ab, casanova_2016_aa, casanova_2018_aa, jose_2014_aa, jose_2020_aa} implied that we could reasonably approximate
 their results by accreting a hydrogen-rich (solar abundances) layer and then switch to a mixed composition once the TNR was underway and convection had begun.
A similar technique has already been used by \cite{jose_2007_aa} who explored a variety of time scales for mixing the WD material into the accreted layers, once convection was
underway, and found that using short time scales was warranted.   Our initial studies, using a similar technique, suggested that accretion with mixing of WD with solar material onto CO WDs,  resulted in CNe explosions where the WD was growing in mass \citep{starrfield_2017_ab, starrfield_2018_aa, starrfield_2018_ab}.  

Therefore, we have now used NOVA to study the consequences of TNRs on WDs of various masses using three different compositions \citep{starrfield_2019_aa}.  In all cases we find that more mass is
accreted than ejected and, therefore, the WD is growing in mass.  We now switch to a mixed composition once the TNR is ongoing and a major fraction of the accreted material is 
convective.  This technique provides a range of model outcomes that are more compatible with observed CNe physical parameters 
reported in the literature.  The simulations with 25\% WD matter and 75\% solar matter appear to fit the observations
somewhat better than those with 50\% WD matter and 50\% solar matter.  

Nevertheless, NOVA  is able to only follow one
 outburst and reaching close to the Chandrasekhar Limit requires many such cycles of accretion-TNR-ejection-accretion.
 While this has yet to be done with either CO or ONe enriched material (this may have been done in the study of
 \citep{rukeya_2017_aa} but they only reported their ejected mass not the accreted mass), multi-cycle evolution and the growth in mass of the
 WD has been done with solar accretion studies \citep{starrfield_2014_aa, hillman_2015_ab, hillman_2015_ac, hillman_2016_aa, starrfield_2017_aa}.  
 The multi-cycle studies reported in \cite{starrfield_2014_aa}, \cite{starrfield_2017_aa}, and this work were done with 
 MESA \citep[][and references therein]{paxton_2011_aa, paxton_2013_aa, paxton_2018_aa} while those described by
 \cite{hillman_2015_ab} and \cite{hillman_2016_aa} were done with the code of \cite[][and references therein]{kovetz_2009_aa}. 
 Given these studies with multiple codes, therefore, we feel that our single outburst result implies that the consequences of the
 CN outburst is the growth in mass of the WD under all situations.
 
 Of great importance, some of the ejected isotope abundances in the simulations also fit the isotopic ratios measured for some
 pre-solar grains suggesting that these grains come from CNe ejecta \citep{pepin_2011_aa, bose_2019_aa, bose_2019_ab, iliadis_2018_aa}. 
\cite{pepin_2011_aa} studied the neon and helium abundances in cometary dust and identified grains with probable CNe origins.  \cite{bose_2019_aa, bose_2019_ab} compared the compositions of 30 pre-solar SiC grains with predicted isotopic abundances.
 The simulations with 25\% WD matter and 75\% solar matter and with CO WD masses from 0.8 M$_\odot$ to 1.35 M$_\odot$ 
provided the best fits to the measured isotopic data for four SiC grains.  In addition, one grain matched the 50\% WD and 50\% solar 1.35 M$_\odot$ simulation.  \cite{iliadis_2018_aa} reported on a Monte Carlo technique, that involved the random sampling over the most important nova model parameters: the WD composition; the mixing of the outer WD layers with the accreted material before the explosion; the peak temperature and density; the explosion timescales; and the possible dilution of the ejecta after the outburst.  They identified 18 pre-solar grains with measured isotopic signatures consistent with a CO nova origin, without assuming any dilution of the ejecta.  These results emphasize our contention that CNe ejected matter was
present during the formation of the solar system.

More recently, we have compared the isotope abundances from these CO and ONe nova simulations to the O-anomalous pre-solar dust grains with oxide and silicate chemistries. A smaller fraction of the O-anomalous pre-solar grains in meteorites include those with very large $^{17}$O excesses ($^{17}$O/$^{16}$O $> 4 \times 10^{-3}$) and small to moderate depletions in $^{18}$O. These grains were classified as extreme Group 1 grains. The previous models \citep{jose_2004_aa} that made the case for both CO and ONe novae being suitable sites for extreme Group 1 grains 
\citep{vollmer_2008_aa, nittler_2008_aa, gyngard_2010_aa} failed to explain the exact compositions of the extreme Group 1 grains but instead suggested a mixing between the nova ejecta and the surrounding medium. The 0.8 M$_\odot$ and 1.0 M$_\odot$ WD simulations can ideally explain the entire range of observed oxygen isotope compositions with variable amounts (5 to 80\%) of CN ejecta. These same mixing models, however,  cannot explain the silicon or magnesium grain compositions of the pre-solar grains. Alternatively, the only case that works reasonably well for silicon and magnesium isotope compositions of extreme Group 1 grains is the 0.6 M$_\odot$ CO WD MDTNR simulations. It can explain the small enrichments (up to 100 \textperthousand) observed for $^{30}$Si and up to 1000 \textperthousand ~ enrichments observed for magnesium isotope compositions. However, these simulations cannot explain the oxygen isotope compositions. It underproduces both $^{17}$O and $^{18}$O compared to $^{16}$O. $^{17}$O production by explosive proton burning is common in nova explosions but additional $^{18}$O production is required to explain the extreme Group 1 grain compositions. $^{18}$O is produced by helium-burning that would occur in He novae (e.g., V445 Pup that shows no hydrogen lines). Extreme Group 1 grains can potentially form in He novae but these binary systems need to be modeled. Alternatively, $^{18}$O production may occur in the secondary carbon star, which is subsequently accreted onto the WD and takes part in grain condensation.

\section {Conclusions}
\label{conclusions}

We have described the 
thermonuclear evolution that occurs as a consequence of 
accretion onto WDs assuming all compositions (solar, CO, and ONe), WD masses, and mass accretion rates.
If the SD scenario for the progenitors of SN Ia is valid, then 
the growth of a CO or ONe WD to the Chandrasekhar Limit is required.  This, in turn, requires that
more material remain on a WD after a TNR than is ejected by the TNR.  
The hydrodynamic
simulations of accretion onto WDs show that a TNR
always occurs. We have evolved a broad range in WD mass (0.4 M$_\odot$  to 1.35 M$_\odot$)
with NOVA , assuming a solar composition, and found that the maximum amount of ejected material ($\sim4\%$) occurs 
for the 1.25M$_\odot$  sequences and decreases to $\sim 0.1\%$ for the 0.7M$_\odot$ sequences.  
Therefore, the WDs are growing in mass as a consequence of the accretion of solar material.  Finally, 
the time to the TNR is sufficiently short that Recurrent Novae could occur on a much broader range of WD mass than heretofore
believed. We have also done simulations, using NOVA, of TNRs on both CO \citep{starrfield_2019_aa} and ONe (Starrfield et al. 2020, in preparation) WDs with the same results. 

The simulations done with MESA allowed us to study the effects of repeated TNRs on
WDs of various masses and \.M.  Figure  \ref{fig:fig7} illustrates the main conclusions from this second
part of the study.  It shows the parameter space investigated with MESA in terms of WD mass and accretion rate and that the WD mass is growing for virtually all reasonable WD masses and \.M (in the sense of observed CV accretion rates).  We identify the different regimes of behavior in the simulations reported here and show
that, for a large range of accretion rates, the WDs grow in mass and that it is possible
for a WD to start off with a mass as low as 0.70M$_\odot$ and, given sufficient time,  reach 1.35M$_\odot$ and
higher.   The 1.35M$_\odot$ sequences never become red giants and despite recurrent hydrogen and helium flashes their growth in mass continues through the simulations.  Therefore, once an accreting 
WD reaches a mass of 1.35$_\odot$ it will continue growing in mass for all accretion rates that
were studied and could eventually reach the Chandrasekhar Limit.  We emphasize that the high accretion rates
on lower mass WDs that end with growth to red giant dimensions are larger than observed for
typical CVs. 

We find that the helium flashes in
these simulations are insufficiently powerful to eject mass and offset the mass growth. For the WD simulations that grow in
mass, the timescale for reaching the Chandrasekhar Limit varies by two orders of magnitude.
This timescale depends on the WD  mass and \.M and varies from
 $5.6 \times 10^{6}$yr to $ 4.3 \times 10^{8}$yr.  

Finally, we conclude that the SD scenario is a viable channel for 
progenitors of SN Ia \index{Supernova of Type Ia (SN Ia)}explosions. 
However, continuous accretion at high rates onto lower mass WDs should cause them to
be extremely luminous and this is probably ruled out by
the observations.  The only systems that might be accreting at these high rates are the
Super Soft Binary X-ray Sources 
originally discovered in the LMC. 
These systems are binaries, with luminosities L$_* \sim
10^{37-38}$erg s$^{-1}$ and effective temperatures ranging from $3 - 7
\times 10^5$K.   If the WDs are sufficiently massive, and the accreting material
is not mixing with core material, then these results also suggest that the WDs in these systems are
growing in mass.

\section{Acknowledgements}
We acknowledge useful discussion and encouragement from M. Darnley, E. Aydi, J. Jos\'e,  M. Hernanz, A. Heger, S. Kafka,  L. Izzo, P. Molaro, M. Orio, M. della Valle, A. Shafter and the attendees at EWASS18, COSPAR 2018, and HEAD 2019 for  their comments.  This work was supported in part by NASA under the Astrophysics Theory Program grant 14-ATP14-0007 and the U.S. DOE under Contract No. DE-FG02- 97ER41041. SS acknowledges partial support from NASA, NSF and HST grants to ASU, WRH is supported by the U.S. Department of Energy, Office of Nuclear Physics, and CEW acknowledges support from NASA and NSF.   SS thanks F. Giovanelli for inviting him to give this talk.

\bibliography{references_iliadis,starrfield_master}

\begin{thebibliography}{113}
\expandafter\ifx\csname natexlab\endcsname\relax\def\natexlab#1{#1}\fi

\bibitem[{{Arnett} {et~al.}(2010){Arnett}, {Meakin}, \&
  {Young}}]{arnett_2010_aa}
{Arnett}, D., {Meakin}, C., \& {Young}, P.~A. 2010, \apj, 710, 1619

\bibitem[{{Bloom} {et~al.}(2012){Bloom}, {Kasen}, {Shen}, {Nugent}, {Butler},
  {Graham}, {Howell}, {Kolb}, {Holmes}, {Haswell}, {Burwitz}, {Rodriguez}, \&
  {Sullivan}}]{bloom_2012_aa}
{Bloom}, J.~S., {Kasen}, D., {Shen}, K.~J., {et~al.} 2012, \apjl, 744, L17

\bibitem[{{Bose} \& {Starrfield}(2019{\natexlab{a}})}]{bose_2019_aa}
{Bose}, M., \& {Starrfield}, S. 2019{\natexlab{a}}, \apj, 873, 14

\bibitem[{{Bose} \& {Starrfield}(2019{\natexlab{b}})}]{bose_2019_ab}
---. 2019{\natexlab{b}}, \apj, 873, 14

\bibitem[{{Branch} {et~al.}(1995){Branch}, {Livio}, {Yungelson}, {Boffi}, \&
  {Baron}}]{branch_1995_aa}
{Branch}, D., {Livio}, M., {Yungelson}, L.~R., {Boffi}, F.~R., \& {Baron}, E.
  1995, \pasp, 107, 1019

\bibitem[{{Cao} {et~al.}(2015){Cao}, {Kulkarni}, {Howell}, {Gal-Yam},
  {Kasliwal}, {Valenti}, {Johansson}, {Amanullah}, {Goobar}, {Sollerman},
  {Taddia}, {Horesh}, {Sagiv}, {Cenko}, {Nugent}, {Arcavi}, {Surace},
  {Wo{\'z}niak}, {Moody}, {Rebbapragada}, {Bue}, \& {Gehrels}}]{cao_2015_aa}
{Cao}, Y., {Kulkarni}, S.~R., {Howell}, D.~A., {et~al.} 2015, \nat, 521, 328

\bibitem[{{Casanova} {et~al.}(2010{\natexlab{a}}){Casanova}, {Jos{\'e}},
  {Garc{\'{\i}}a-Berro}, {Calder}, \& {Shore}}]{casanova_2010_aa}
{Casanova}, J., {Jos{\'e}}, J., {Garc{\'{\i}}a-Berro}, E., {Calder}, A., \&
  {Shore}, S.~N. 2010{\natexlab{a}}, \aap, 513, L5+

\bibitem[{{Casanova} {et~al.}(2010{\natexlab{b}}){Casanova}, {Jos{\'e}},
  {Garc{\'{\i}}a-Berro}, {Calder}, \& {Shore}}]{casanova_2010_ab}
---. 2010{\natexlab{b}}, \aap, 513, L5

\bibitem[{{Casanova} {et~al.}(2011{\natexlab{a}}){Casanova}, {Jos{\'e}},
  {Garc{\'{\i}}a-Berro}, {Calder}, \& {Shore}}]{casanova_2011_aa}
---. 2011{\natexlab{a}}, \aap, 527, A5

\bibitem[{{Casanova} {et~al.}(2016){Casanova}, {Jos{\'e}}, {Garc{\'\i}a-Berro},
  \& {Shore}}]{casanova_2016_aa}
{Casanova}, J., {Jos{\'e}}, J., {Garc{\'\i}a-Berro}, E., \& {Shore}, S.~N.
  2016, \aap, 595, A28

\bibitem[{{Casanova} {et~al.}(2011{\natexlab{b}}){Casanova}, {Jos{\'e}},
  {Garc{\'{\i}}a-Berro}, {Shore}, \& {Calder}}]{casanova_2011_ab}
{Casanova}, J., {Jos{\'e}}, J., {Garc{\'{\i}}a-Berro}, E., {Shore}, S.~N., \&
  {Calder}, A.~C. 2011{\natexlab{b}}, \nat, 478, 490

\bibitem[{{Casanova} {et~al.}(2018){Casanova}, {Jos{\'e}}, \&
  {Shore}}]{casanova_2018_aa}
{Casanova}, J., {Jos{\'e}}, J., \& {Shore}, S.~N. 2018, \aap, 619, A121

\bibitem[{{Cassisi} {et~al.}(2007){Cassisi}, {Potekhin}, {Pietrinferni},
  {Catelan}, \& {Salaris}}]{cassisi_2007_aa}
{Cassisi}, S., {Potekhin}, A.~Y., {Pietrinferni}, A., {Catelan}, M., \&
  {Salaris}, M. 2007, \apj, 661, 1094

\bibitem[{{Chomiuk} {et~al.}(2012){Chomiuk}, {Soderberg}, {Moe}, {Chevalier},
  {Rupen}, {Badenes}, {Margutti}, {Fransson}, {Fong}, \&
  {Dittmann}}]{chomiuk_2012_aa}
{Chomiuk}, L., {Soderberg}, A.~M., {Moe}, M., {et~al.} 2012, \apj, 750, 164

\bibitem[{{Copperwheat} {et~al.}(2010){Copperwheat}, {Marsh}, {Dhillon},
  {Littlefair}, {Hickman}, {G{\"a}nsicke}, \&
  {Southworth}}]{copperwheat_2010_aa}
{Copperwheat}, C.~M., {Marsh}, T.~R., {Dhillon}, V.~S., {et~al.} 2010, \mnras,
  402, 1824

\bibitem[{{Cowley} {et~al.}(1998){Cowley}, {Schmidtke}, {Crampton}, \&
  {Hutchings}}]{cowley_1998_aa}
{Cowley}, A.~P., {Schmidtke}, P.~C., {Crampton}, D., \& {Hutchings}, J.~B.
  1998, \apj, 504, 854

\bibitem[{{Darnley} {et~al.}(2016){Darnley}, {Henze}, {Bode}, {Hachisu},
  {Hernanz}, {Hornoch}, {Hounsell}, {Kato}, {Ness}, {Osborne}, {Page},
  {Ribeiro}, {Rodr{\'\i}guez-Gil}, {Shafter}, {Shara}, {Steele}, {Williams},
  {Arai}, {Arcavi}, {Barsukova}, {Boumis}, {Chen}, {Fabrika}, {Figueira},
  {Gao}, {Gehrels}, {Godon}, {Goranskij}, {Harman}, {Hartmann}, {Hosseinzadeh},
  {Horst}, {Itagaki}, {Jos{\'e}}, {Kabashima}, {Kaur}, {Kawai}, {Kennea},
  {Kiyota}, {Ku{\v{c}}{\'a}kov{\'a}}, {Lau}, {Maehara}, {Naito}, {Nakajima},
  {Nishiyama}, {O'Brien}, {Quimby}, {Sala}, {Sano}, {Sion}, {Valeev},
  {Watanabe}, {Watanabe}, {Williams}, \& {Xu}}]{darnley_2016_aa}
{Darnley}, M.~J., {Henze}, M., {Bode}, M.~F., {et~al.} 2016, \apj, 833, 149

\bibitem[{{Darnley} {et~al.}(2017{\natexlab{a}}){Darnley}, {Hounsell}, {Godon},
  {Perley}, {Henze}, {Kuin}, {Williams}, {Williams}, {Bode}, {Harman},
  {Hornoch}, {Link}, {Ness}, {Ribeiro}, {Sion}, {Shafter}, \&
  {Shara}}]{darnley_2017_ab}
{Darnley}, M.~J., {Hounsell}, R., {Godon}, P., {et~al.} 2017{\natexlab{a}},
  \apj, 849, 96

\bibitem[{{Darnley} {et~al.}(2017{\natexlab{b}}){Darnley}, {Hounsell}, {Godon},
  {Perley}, {Henze}, {Kuin}, {Williams}, {Williams}, {Bode}, {Harman},
  {Hornoch}, {Link}, {Ness}, {Ribeiro}, {Sion}, {Shafter}, \&
  {Shara}}]{darnley_2017_aa}
---. 2017{\natexlab{b}}, \apj, 847, 35

\bibitem[{{Darnley} {et~al.}(2019){Darnley}, {Hounsell}, {O'Brien}, {Henze},
  {Rodr{\'{\i}}guez-Gil}, {Shafter}, {Shara}, {Vaytet}, {Bode}, {Ciardullo},
  {Davis}, {Galera-Rosillo}, {Harman}, {Harvey}, {Healy}, {Ness}, {Ribeiro}, \&
  {Williams}}]{darnley_2019_aa}
{Darnley}, M.~J., {Hounsell}, R., {O'Brien}, T.~J., {et~al.} 2019, \nat, 565,
  460

\bibitem[{{Echevarr{\'{\i}}a} {et~al.}(2007){Echevarr{\'{\i}}a}, {de la
  Fuente}, \& {Costero}}]{Echevarria_2007_aa}
{Echevarr{\'{\i}}a}, J., {de la Fuente}, E., \& {Costero}, R. 2007, \aj, 134,
  262

\bibitem[{{Edwards} {et~al.}(2012){Edwards}, {Pagnotta}, \&
  {Schaefer}}]{edwardspag_2012_aa}
{Edwards}, Z.~I., {Pagnotta}, A., \& {Schaefer}, B.~E. 2012, \apjl, 747, L19

\bibitem[{{Ferguson} {et~al.}(2005){Ferguson}, {Alexander}, {Allard}, {Barman},
  {Bodnarik}, {Hauschildt}, {Heffner-Wong}, \& {Tamanai}}]{ferguson_2005_aa}
{Ferguson}, J.~W., {Alexander}, D.~R., {Allard}, F., {et~al.} 2005, \apj, 623,
  585

\bibitem[{{Fujimoto}(1982{\natexlab{a}})}]{fujimoto_1982_ab}
{Fujimoto}, M.~Y. 1982{\natexlab{a}}, \apj, 257, 767

\bibitem[{{Fujimoto}(1982{\natexlab{b}})}]{fujimoto_1982_aa}
---. 1982{\natexlab{b}}, \apj, 257, 752

\bibitem[{{Gilfanov} \& {Bogd{\'a}n}(2010)}]{gilfanov_2010_aa}
{Gilfanov}, M., \& {Bogd{\'a}n}, {\'A}. 2010, \nat, 463, 924

\bibitem[{{Graur} {et~al.}(2014){Graur}, {Maoz}, \& {Shara}}]{graur_2014_aa}
{Graur}, O., {Maoz}, D., \& {Shara}, M.~M. 2014, \mnras, 442, L28

\bibitem[{{Graur} \& {Woods}(2019)}]{graur_2019_aa}
{Graur}, O., \& {Woods}, T.~E. 2019, \mnras, 484, L79

\bibitem[{{Gyngard} {et~al.}(2010){Gyngard}, {Nittler}, {Zinner}, \&
  {Jose}}]{gyngard_2010_aa}
{Gyngard}, F., {Nittler}, L., {Zinner}, E., \& {Jose}, J. 2010, in Nuclei in
  the Cosmos, 141

\bibitem[{{Henden} {et~al.}(2001){Henden}, {Wagner}, \&
  {Starrfield}}]{henden_2001_aa}
{Henden}, A.~A., {Wagner}, R.~M., \& {Starrfield}, S.~G. 2001, \iaucirc, 7730,
  1

\bibitem[{{Henze} {et~al.}(2015){Henze}, {Darnley}, {Kabashima}, {Nishiyama},
  {Itagaki}, \& {Gao}}]{henze_2015_aa}
{Henze}, M., {Darnley}, M.~J., {Kabashima}, F., {et~al.} 2015, \aap, 582, L8

\bibitem[{{Henze} {et~al.}(2018){Henze}, {Darnley}, {Williams}, {Kato},
  {Hachisu}, {Anupama}, {Arai}, {Boyd}, {Burke}, {Ciardullo}, {Chinetti},
  {Cook}, {Cook}, {Erdman}, {Gao}, {Harris}, {Hartmann}, {Hornoch}, {Horst},
  {Hounsell}, {Husar}, {Itagaki}, {Kabashima}, {Kafka}, {Kaur}, {Kiyota},
  {Kojiguchi}, {Ku{\v c}{\'a}kov{\'a}}, {Kuramoto}, {Maehara}, {Mantero},
  {Masci}, {Matsumoto}, {Naito}, {Ness}, {Nishiyama}, {Oksanen}, {Osborne},
  {Page}, {Paunzen}, {Pavana}, {Pickard}, {Prieto-Arranz},
  {Rodr{\'{\i}}guez-Gil}, {Sala}, {Sano}, {Shafter}, {Sugiura}, {Tan},
  {Tordai}, {Vra{\v s}til}, {Wagner}, {Watanabe}, {Williams}, {Bode}, {Bruno},
  {Buchheim}, {Crawford}, {Goff}, {Hernanz}, {Igarashi}, {Jos{\'e}}, {Motta},
  {O'Brien}, {Oswalt}, {Poyner}, {Ribeiro}, {Sabo}, {Shara}, {Shears},
  {Starkey}, {Starrfield}, \& {Woodward}}]{henze_2018_aa}
{Henze}, M., {Darnley}, M.~J., {Williams}, S.~C., {et~al.} 2018, \apj, 857, 68

\bibitem[{{Hillebrandt} \& {Leibundgut}(2003)}]{hillebrandt_2003_aa}
{Hillebrandt}, W., \& {Leibundgut}, B., eds. 2003, {From twilight to highlight
  : the physics of supernovae : proceedings of the ESO/MPA/MPE workshop held at
  Garching, Germany, 29-31 July 2002}

\bibitem[{{Hillebrandt} \& {Niemeyer}(2000)}]{hillebrandt_2000_aa}
{Hillebrandt}, W., \& {Niemeyer}, J.~C. 2000, \araa, 38, 191

\bibitem[{{Hillman} {et~al.}(2015{\natexlab{a}}){Hillman}, {Prialnik},
  {Kovetz}, \& {Shara}}]{hillman_2015_ab}
{Hillman}, Y., {Prialnik}, D., {Kovetz}, A., \& {Shara}, M.~M.
  2015{\natexlab{a}}, in Astronomical Society of the Pacific Conference Series,
  Vol. 493, 19th European Workshop on White Dwarfs, ed. P.~{Dufour},
  P.~{Bergeron}, \& G.~{Fontaine}, 553

\bibitem[{{Hillman} {et~al.}(2015{\natexlab{b}}){Hillman}, {Prialnik},
  {Kovetz}, \& {Shara}}]{hillman_2015_ac}
{Hillman}, Y., {Prialnik}, D., {Kovetz}, A., \& {Shara}, M.~M.
  2015{\natexlab{b}}, \mnras, 446, 1924

\bibitem[{{Hillman} {et~al.}(2016){Hillman}, {Prialnik}, {Kovetz}, \&
  {Shara}}]{hillman_2016_aa}
---. 2016, \apj, 819, 168

\bibitem[{{Hix} \& {Thielemann}(1999)}]{hix_1999_aa}
{Hix}, W.~R., \& {Thielemann}, F.-K. 1999, \apj, 511, 862

\bibitem[{{Howell}(2010)}]{howell_2010_ab}
{Howell}, D.~A. 2010, \nat, 463, 35

\bibitem[{{Howell}(2011)}]{howell_2011_aa}
---. 2011, Nature Communications, 2

\bibitem[{{Howell} {et~al.}(2009){Howell}, {Conley}, {Della Valle}, {Nugent},
  {Perlmutter}, {Marion}, {Krisciunas}, {Badenes}, {Mazzali}, {Aldering},
  {Antilogus}, {Baron}, {Becker}, {Baltay}, {Benetti}, {Blondin}, {Branch},
  {Brown}, {Deustua}, {Ealet}, {Ellis}, {Fouchez}, {Freedman}, {Gal-Yam},
  {Jha}, {Kasen}, {Kessler}, {Kim}, {Leonard}, {Li}, {Livio}, {Maoz},
  {Mannucci}, {Matheson}, {Neill}, {Nomoto}, {Panagia}, {Perrett}, {Phillips},
  {Poznanski}, {Quimby}, {Rest}, {Riess}, {Sako}, {Soderberg}, {Strolger},
  {Thomas}, {Turatto}, {van Dyk}, \& {Wood-Vasey}}]{howell_2009_ab}
{Howell}, D.~A., {Conley}, A., {Della Valle}, M., {et~al.} 2009, ArXiv
  e-prints: White Paper Submitted to the Astro 2010 Decadel Survey

\bibitem[{{Iben}(1991)}]{iben_1991_aa}
{Iben}, Jr., I. 1991, \apjs, 76, 55

\bibitem[{{Iben} {et~al.}(1997){Iben}, {Ritossa}, \&
  {Garc{\'\i}a-Berro}}]{iben_1997_aa}
{Iben}, Icko, J., {Ritossa}, C., \& {Garc{\'\i}a-Berro}, E. 1997, \apj, 489,
  772

\bibitem[{{Iglesias} \& {Rogers}(1996)}]{iglesias_1996_aa}
{Iglesias}, C.~A., \& {Rogers}, F.~J. 1996, \apj, 464, 943

\bibitem[{{Iliadis} {et~al.}(2018){Iliadis}, {Downen}, {Jos{\'e}}, {Nittler},
  \& {Starrfield}}]{iliadis_2018_aa}
{Iliadis}, C., {Downen}, L.~N., {Jos{\'e}}, J., {Nittler}, L.~R., \&
  {Starrfield}, S. 2018, \apj, 855, 76

\bibitem[{{Jos{\'e}}(2014)}]{jose_2014_aa}
{Jos{\'e}}, J. 2014, in Astronomical Society of the Pacific Conference Series,
  Vol. 490, Stella Novae: Past and Future Decades, ed. P.~A. {Woudt} \&
  V.~A.~R.~M. {Ribeiro}, 275

\bibitem[{{Jos{\'e}} {et~al.}(2007){Jos{\'e}}, {Garc{\'{\i}}a-Berro},
  {Hernanz}, \& {Gil-Pons}}]{jose_2007_aa}
{Jos{\'e}}, J., {Garc{\'{\i}}a-Berro}, E., {Hernanz}, M., \& {Gil-Pons}, P.
  2007, \apjl, 662, L103

\bibitem[{{Jos{\'e}} {et~al.}(2004){Jos{\'e}}, {Hernanz}, {Amari}, {Lodders},
  \& {Zinner}}]{jose_2004_aa}
{Jos{\'e}}, J., {Hernanz}, M., {Amari}, S., {Lodders}, K., \& {Zinner}, E.
  2004, \apj, 612, 414

\bibitem[{{Jos{\'e}} {et~al.}(2020){Jos{\'e}}, {Shore}, \&
  {Casanova}}]{jose_2020_aa}
{Jos{\'e}}, J., {Shore}, S.~N., \& {Casanova}, J. 2020, \aap, 634, A5

\bibitem[{{Kahabka} \& {van den Heuvel}(1997)}]{kahabka_1997_aa}
{Kahabka}, P., \& {van den Heuvel}, E.~P.~J. 1997, \araa, 35, 69

\bibitem[{Kasen {et~al.}(2009)Kasen, R\"{o}pke, \& Woosley}]{kasen_2009_aa}
Kasen, D., R\"{o}pke, F.~K., \& Woosley, S.~E. 2009, Nature, 460, 869

\bibitem[{{Khokhlov}(1991)}]{khokhlov_1991_aa}
{Khokhlov}, A.~M. 1991, \aap, 245, 114

\bibitem[{{Kovetz} {et~al.}(2009){Kovetz}, {Yaron}, \&
  {Prialnik}}]{kovetz_2009_aa}
{Kovetz}, A., {Yaron}, O., \& {Prialnik}, D. 2009, \mnras, 395, 1857

\bibitem[{{Kuuttila} {et~al.}(2019){Kuuttila}, {Gilfanov}, {Seitenzahl},
  {Woods}, \& {Vogt}}]{kuuttila_2019_aa}
{Kuuttila}, J., {Gilfanov}, M., {Seitenzahl}, I.~R., {Woods}, T.~E., \& {Vogt},
  F.~P.~A. 2019, \mnras, 484, 1317

\bibitem[{{Leibundgut}(2000)}]{leibundgut_2000_aa}
{Leibundgut}, B. 2000, \aapr, 10, 179

\bibitem[{{Leibundgut}(2001)}]{leibundgut_2001_aa}
---. 2001, \araa, 39, 67

\bibitem[{{Li} {et~al.}(2011){Li}, {Bloom}, {Podsiadlowski}, {Miller}, {Cenko},
  {Jha}, {Sullivan}, {Howell}, {Nugent}, {Butler}, {Ofek}, {Kasliwal},
  {Richards}, {Stockton}, {Shih}, {Bildsten}, {Shara}, {Bibby}, {Filippenko},
  {Ganeshalingam}, {Silverman}, {Kulkarni}, {Law}, {Poznanski}, {Quimby},
  {McCully}, {Patel}, {Maguire}, \& {Shen}}]{weidongli_2011_aa}
{Li}, W., {Bloom}, J.~S., {Podsiadlowski}, P., {et~al.} 2011, \nat, 480, 348

\bibitem[{{Lundqvist} {et~al.}(2015){Lundqvist}, {Nyholm}, {Taddia},
  {Sollerman}, {Johansson}, {Kozma}, {Lundqvist}, {Fransson}, {Garnavich},
  {Kromer}, {Shappee}, \& {Goobar}}]{lundqvist_2015_aa}
{Lundqvist}, P., {Nyholm}, A., {Taddia}, F., {et~al.} 2015, \aap, 577, A39

\bibitem[{{Lyke} {et~al.}(2001){Lyke}, {Woodward}, {Gehrz}, {Wagner},
  {Starrfield}, {Schwarz}, \& {Foltz}}]{lyke_2001_aa}
{Lyke}, J.~E., {Woodward}, C.~E., {Gehrz}, R.~D., {et~al.} 2001, in Bulletin of
  the American Astronomical Society, Vol.~33, American Astronomical Society
  Meeting Abstracts \#198, 803

\bibitem[{{Maoz} {et~al.}(2014){Maoz}, {Mannucci}, \&
  {Nelemans}}]{maoz_2014_aa}
{Maoz}, D., {Mannucci}, F., \& {Nelemans}, G. 2014, \araa, 52, 107

\bibitem[{{Newsham} {et~al.}(2014){Newsham}, {Starrfield}, \&
  {Timmes}}]{newsham_2014_aa}
{Newsham}, G., {Starrfield}, S., \& {Timmes}, F.~X. 2014, in Astronomical
  Society of the Pacific Conference Series, Vol. 490, Stella Novae: Past and
  Future Decades, ed. P.~A. {Woudt} \& V.~A.~R.~M. {Ribeiro}, 287

\bibitem[{{Nittler} {et~al.}(2008){Nittler}, {Alexander}, {Gallino}, {Hoppe},
  {Nguyen}, {Stadermann}, \& {Zinner}}]{nittler_2008_aa}
{Nittler}, L.~R., {Alexander}, C. M.~O., {Gallino}, R., {et~al.} 2008, \apj,
  682, 1450

\bibitem[{{Nomoto}(1982)}]{nomoto_1982_aa}
{Nomoto}, K. 1982, \apj, 253, 798

\bibitem[{{Nomoto} {et~al.}(2003){Nomoto}, {Uenishi}, {Kobayashi}, {Umeda},
  {Ohkubo}, {Hachisu}, \& {Kato}}]{nomoto_2003_aa}
{Nomoto}, K., {Uenishi}, T., {Kobayashi}, C., {et~al.} 2003, in From Twilight
  to Highlight: The Physics of Supernovae, ed. W.~{Hillebrandt} \&
  B.~{Leibundgut}, 115--+

\bibitem[{{Nugent} {et~al.}(2011){Nugent}, {Sullivan}, {Cenko}, {Thomas},
  {Kasen}, {Howell}, {Bersier}, {Bloom}, {Kulkarni}, {Kandrashoff},
  {Filippenko}, {Silverman}, {Marcy}, {Howard}, {Isaacson}, {Maguire},
  {Suzuki}, {Tarlton}, {Pan}, {Bildsten}, {Fulton}, {Parrent}, {Sand},
  {Podsiadlowski}, {Bianco}, {Dilday}, {Graham}, {Lyman}, {James}, {Kasliwal},
  {Law}, {Quimby}, {Hook}, {Walker}, {Mazzali}, {Pian}, {Ofek}, {Gal-Yam}, \&
  {Poznanski}}]{pnugent11}
{Nugent}, P.~E., {Sullivan}, M., {Cenko}, S.~B., {et~al.} 2011, \nat, 480, 344

\bibitem[{{Olling} {et~al.}(2015){Olling}, {Mushotzky}, {Shaya}, {Rest},
  {Garnavich}, {Tucker}, {Kasen}, {Margheim}, \& {Filippenko}}]{olling_2015_aa}
{Olling}, R.~P., {Mushotzky}, R., {Shaya}, E.~J., {et~al.} 2015, \nat, 521, 332

\bibitem[{{Paxton} {et~al.}(2011){Paxton}, {Bildsten}, {Dotter}, {Herwig},
  {Lesaffre}, \& {Timmes}}]{paxton_2011_aa}
{Paxton}, B., {Bildsten}, L., {Dotter}, A., {et~al.} 2011, \apjs, 192, 3

\bibitem[{{Paxton} {et~al.}(2013){Paxton}, {Cantiello}, {Arras}, {Bildsten},
  {Brown}, {Dotter}, {Mankovich}, {Montgomery}, {Stello}, {Timmes}, \&
  {Townsend}}]{paxton_2013_aa}
{Paxton}, B., {Cantiello}, M., {Arras}, P., {et~al.} 2013, \apjs, 208, 4

\bibitem[{{Paxton} {et~al.}(2015){Paxton}, {Marchant}, {Schwab}, {Bauer},
  {Bildsten}, {Cantiello}, {Dessart}, {Farmer}, {Hu}, {Langer}, {Townsend},
  {Townsley}, \& {Timmes}}]{paxton_2015_aa}
{Paxton}, B., {Marchant}, P., {Schwab}, J., {et~al.} 2015, \apjs, 220, 15

\bibitem[{{Paxton} {et~al.}(2018){Paxton}, {Schwab}, {Bauer}, {Bildsten},
  {Blinnikov}, {Duffell}, {Farmer}, {Goldberg}, {Marchant}, {Sorokina},
  {Thoul}, {Townsend}, \& {Timmes}}]{paxton_2018_aa}
{Paxton}, B., {Schwab}, J., {Bauer}, E.~B., {et~al.} 2018, \apjs, 234, 34

\bibitem[{{Paxton} {et~al.}(2019){Paxton}, {Smolec}, {Schwab}, {Gautschy},
  {Bildsten}, {Cantiello}, {Dotter}, {Farmer}, {Goldberg}, {Jermyn}, {Kanbur},
  {Marchant}, {Thoul}, {Townsend}, {Wolf}, {Zhang}, \&
  {Timmes}}]{paxton_2019_aa}
{Paxton}, B., {Smolec}, R., {Schwab}, J., {et~al.} 2019, \apjs, 243, 10

\bibitem[{{Pepin} {et~al.}(2011){Pepin}, {Palma}, {Gehrz}, \&
  {Starrfield}}]{pepin_2011_aa}
{Pepin}, R.~O., {Palma}, R.~L., {Gehrz}, R.~D., \& {Starrfield}, S. 2011, \apj,
  742, 86

\bibitem[{{Potekhin} \& {Chabrier}(2010)}]{potekhin_2010_aa}
{Potekhin}, A.~Y., \& {Chabrier}, G. 2010, Contributions to Plasma Physics, 50,
  82

\bibitem[{{Ritossa} {et~al.}(1996){Ritossa}, {Garcia-Berro}, \&
  {Iben}}]{ritossa_1996_aa}
{Ritossa}, C., {Garcia-Berro}, E., \& {Iben}, Jr., I. 1996, \apj, 460, 489

\bibitem[{{Rogers} \& {Nayfonov}(2002)}]{rogers_2002_aa}
{Rogers}, F.~J., \& {Nayfonov}, A. 2002, \apj, 576, 1064

\bibitem[{{Ruiter}(2020)}]{ruiter_2020_aa}
{Ruiter}, A.~J. 2020, arXiv e-prints, arXiv:2001.02947

\bibitem[{{Ruiz-Lapuente}(2014)}]{ruiz_2014_aa}
{Ruiz-Lapuente}, P. 2014, New Astronomy Reviews, 62, 15

\bibitem[{{Rukeya} {et~al.}(2017){Rukeya}, {L{\"u}}, {Wang}, \&
  {Zhu}}]{rukeya_2017_aa}
{Rukeya}, R., {L{\"u}}, G., {Wang}, Z., \& {Zhu}, C. 2017, \pasp, 129, 074201

\bibitem[{{Sallaska} {et~al.}(2013){Sallaska}, {Iliadis}, {Champagne},
  {Goriely}, {Starrfield}, \& {Timmes}}]{sallaska_2013_aa}
{Sallaska}, A.~L., {Iliadis}, C., {Champagne}, A.~E., {et~al.} 2013, \apjs,
  207, 18

\bibitem[{{Saumon} {et~al.}(1995){Saumon}, {Chabrier}, \& {van
  Horn}}]{saumon_1995_aa}
{Saumon}, D., {Chabrier}, G., \& {van Horn}, H.~M. 1995, \apjs, 99, 713

\bibitem[{{Schaefer} \& {Pagnotta}(2012)}]{schaeferpag_2012_aa}
{Schaefer}, B.~E., \& {Pagnotta}, A. 2012, \nat, 481, 164

\bibitem[{{Schwarz} {et~al.}(2001){Schwarz}, {Shore}, {Starrfield},
  {Hauschildt}, {Della Valle}, \& {Baron}}]{schwarz_2001_aa}
{Schwarz}, G.~J., {Shore}, S.~N., {Starrfield}, S., {et~al.} 2001, \mnras, 320,
  103

\bibitem[{{Schwarzschild} \& {H{\"a}rm}(1965)}]{schwarzschild_1965_aa}
{Schwarzschild}, M., \& {H{\"a}rm}, R. 1965, \apj, 142, 855

\bibitem[{{Selvelli} \& {Gilmozzi}(2019)}]{selvelli_2019_aa}
{Selvelli}, P., \& {Gilmozzi}, R. 2019, \aap, 622, A186

\bibitem[{{Shafter}(1983)}]{Shafter_1983_aa}
{Shafter}, A.~W. 1983, PhD thesis, California Univ., Los Angeles.

\bibitem[{{Shara} {et~al.}(2018){Shara}, {Prialnik}, {Hillman}, \&
  {Kovetz}}]{shara_2018_ab}
{Shara}, M.~M., {Prialnik}, D., {Hillman}, Y., \& {Kovetz}, A. 2018, \apj, 860,
  110

\bibitem[{{Shaviv}(2002)}]{shavivnj_2002_aa}
{Shaviv}, N.~J. 2002, in American Institute of Physics Conference Series, Vol.
  637, Classical Nova Explosions, ed. M.~{Hernanz} \& J.~{Jos{\'e}}, 259--265

\bibitem[{{Sion}(1986)}]{sion_1986_aa}
{Sion}, E.~M. 1986, Publications of the Astronomical Society of the Pacific,
  98, 821

\bibitem[{{Sion} {et~al.}(2010){Sion}, {Godon}, {Myzcka}, \&
  {Blair}}]{Sion_2010_SSCyg_aa}
{Sion}, E.~M., {Godon}, P., {Myzcka}, J., \& {Blair}, W.~P. 2010, \apjl, 716,
  L157

\bibitem[{{Sion} {et~al.}(2019){Sion}, {Wilson}, {Godon}, {Starrfield},
  {Williams}, \& {Darnley}}]{sion_2019_aa}
{Sion}, E.~M., {Wilson}, R.~E., {Godon}, P., {et~al.} 2019, \apj, 872, 68

\bibitem[{{Starrfield}(2014)}]{starrfield_2014_aa}
{Starrfield}, S. 2014, AIP Advances, 4, 041007

\bibitem[{{Starrfield}(2017)}]{starrfield_2017_aa}
---. 2017, {Evolution of Accreting White Dwarfs to the Thermonuclear Runaway},
  ed. A.~W. {Alsabti} \& P.~{Murdin} (Springer Berlin / Heidelberg), 1211

\bibitem[{{Starrfield} {et~al.}(2017){Starrfield}, {Bose}, {Iliadis}, {Hix},
  {Wagner}, {Woodward}, {Jose}, \& {Hernanz}}]{starrfield_2017_ab}
{Starrfield}, S., {Bose}, M., {Iliadis}, C., {et~al.} 2017, in The Golden Age
  of Cataclysmic Variables and Related Objects IV, 66

\bibitem[{{Starrfield} {et~al.}(2018{\natexlab{a}}){Starrfield}, {Bose},
  {Iliadis}, {Hix}, {Wagner}, {Woodward}, {Jose'}, \&
  {Hernanz}}]{starrfield_2018_aa}
{Starrfield}, S., {Bose}, M., {Iliadis}, C., {et~al.} 2018{\natexlab{a}}, in
  American Astronomical Society Meeting Abstracts, Vol. 231, American
  Astronomical Society Meeting Abstracts \#231, 358.11

\bibitem[{{Starrfield} {et~al.}(2019){Starrfield}, {Bose}, {Iliadis}, {Hix},
  {Woodward}, \& {Wagner}}]{starrfield_2019_aa}
{Starrfield}, S., {Bose}, M., {Iliadis}, C., {et~al.} 2019, arXiv e-prints,
  arXiv:1910.00575

\bibitem[{{Starrfield} {et~al.}(2016){Starrfield}, {Iliadis}, \&
  {Hix}}]{starrfield_2016_aa}
{Starrfield}, S., {Iliadis}, C., \& {Hix}, W.~R. 2016, \pasp, 128, 051001

\bibitem[{{Starrfield} {et~al.}(2009){Starrfield}, {Iliadis}, {Hix}, {Timmes},
  \& {Sparks}}]{starrfieldpep09}
{Starrfield}, S., {Iliadis}, C., {Hix}, W.~R., {Timmes}, F.~X., \& {Sparks},
  W.~M. 2009, \apj, 692, 1532

\bibitem[{{Starrfield} {et~al.}(2012){Starrfield}, {Iliadis}, {Timmes}, {Hix},
  {Arnett}, {Meakin}, \& {Sparks}}]{starrfield_2012_basi}
{Starrfield}, S., {Iliadis}, C., {Timmes}, F.~X., {et~al.} 2012, Bulletin of
  the Astronomical Society of India, 40, 419

\bibitem[{{Starrfield} {et~al.}(1998){Starrfield}, {Truran}, {Wiescher}, \&
  {Sparks}}]{starrfield_1998_aa}
{Starrfield}, S., {Truran}, J.~W., {Wiescher}, M.~C., \& {Sparks}, W.~M. 1998,
  \mnras, 296, 502

\bibitem[{{Starrfield} {et~al.}(2018{\natexlab{b}}){Starrfield}, {Bose},
  {Iliadis}, {Hix}, {Woodward}, {Wagner}, {Jos{\'e}}, {Hernanz}, \&
  {Feng}}]{starrfield_2018_ab}
{Starrfield}, S., {Bose}, M., {Iliadis}, C., {et~al.} 2018{\natexlab{b}}, in
  American Astronomical Society Meeting Abstracts, Vol. 232, American
  Astronomical Society Meeting Abstracts \#232, 320.04

\bibitem[{{Timmes} \& {Arnett}(1999)}]{timmes_1999_aa}
{Timmes}, F.~X., \& {Arnett}, D. 1999, \apjs, 125, 277

\bibitem[{{Timmes} \& {Swesty}(2000)}]{timmes_2000_ab}
{Timmes}, F.~X., \& {Swesty}, F.~D. 2000, \apjs, 126, 501

\bibitem[{{Tomov} {et~al.}(2015){Tomov}, {Swierczynski}, {Mikolajewski}, \&
  {Ilkiewicz}}]{tomov_2015_aa}
{Tomov}, T., {Swierczynski}, E., {Mikolajewski}, M., \& {Ilkiewicz}, K. 2015,
  \aap, 576, A119

\bibitem[{{Tr{\"u}mper} {et~al.}(1991){Tr{\"u}mper}, {Hasinger}, {Aschenbach},
  {Br{\"a}uninger}, {Briel}, {Burkert}, {Fink}, {Pfeffermann}, {Pietsch},
  {Predehl}, {Schmitt}, {Voges}, {Zimmermann}, \&
  {Beuermann}}]{Trumper_1991_aa}
{Tr{\"u}mper}, J., {Hasinger}, G., {Aschenbach}, B., {et~al.} 1991, \nat, 349,
  579

\bibitem[{{van den Heuvel} {et~al.}(1992){van den Heuvel}, {Bhattacharya},
  {Nomoto}, \& {Rappaport}}]{vandenheuvel_1992_aa}
{van den Heuvel}, E.~P.~J., {Bhattacharya}, D., {Nomoto}, K., \& {Rappaport},
  S.~A. 1992, \aap, 262, 97

\bibitem[{{Vollmer} {et~al.}(2008){Vollmer}, {Hoppe}, \&
  {Brenker}}]{vollmer_2008_aa}
{Vollmer}, C., {Hoppe}, P., \& {Brenker}, F.~E. 2008, \apj, 684, 611

\bibitem[{{Wagner} {et~al.}(2001{\natexlab{a}}){Wagner}, {Schwarz}, \&
  {Starrfield}}]{wagner_2001_aa}
{Wagner}, R.~M., {Schwarz}, G., \& {Starrfield}, S.~G. 2001{\natexlab{a}},
  \iaucirc, 7571, 1

\bibitem[{{Wagner} {et~al.}(2001{\natexlab{b}}){Wagner}, {Schwarz},
  {Starrfield}, {Foltz}, {Howell}, \& {Szkody}}]{wagner_2001_ab}
{Wagner}, R.~M., {Schwarz}, G., {Starrfield}, S.~G., {et~al.}
  2001{\natexlab{b}}, \iaucirc, 7717, 2

\bibitem[{{White} {et~al.}(2015){White}, {Kasliwal}, {Nugent}, {Gal-Yam},
  {Howell}, {Sullivan}, {Goobar}, {Piro}, {Bloom}, {Kulkarni}, {Laher},
  {Masci}, {Ofek}, {Surace}, {Ben-Ami}, {Cao}, {Cenko}, {Hook}, {J{\"o}nsson},
  {Matheson}, {Sternberg}, {Quimby}, \& {Yaron}}]{white_2015_aa}
{White}, C.~J., {Kasliwal}, M.~M., {Nugent}, P.~E., {et~al.} 2015, \apj, 799,
  52

\bibitem[{{Woosley} \& {Kasen}(2011)}]{woosley_kasen_11_a}
{Woosley}, S.~E., \& {Kasen}, D. 2011, \apj, 734, 38

\bibitem[{{Woudt} \& {Steeghs}(2005)}]{woudt_2005_aa}
{Woudt}, P.~A., \& {Steeghs}, D. 2005, in American Institute of Physics
  Conference Series, Vol. 797, Interacting Binaries: Accretion, Evolution, and
  Outcomes, ed. L.~{Burderi}, L.~A. {Antonelli}, F.~{D'Antona}, T.~{di Salvo},
  G.~L. {Israel}, L.~{Piersanti}, A.~{Tornamb{\`e}}, \& O.~{Straniero},
  647--650

\bibitem[{{Woudt} {et~al.}(2009){Woudt}, {Steeghs}, {Karovska}, {Warner},
  {Groot}, {Nelemans}, {Roelofs}, {Marsh}, {Nagayama}, {Smits}, \&
  {O'Brien}}]{woudt_2009_aa}
{Woudt}, P.~A., {Steeghs}, D., {Karovska}, M., {et~al.} 2009, \apj, 706, 738

\bibitem[{{Yoon} {et~al.}(2004){Yoon}, {Langer}, \& {van der
  Sluys}}]{yoon_2004_aa}
{Yoon}, S.-C., {Langer}, N., \& {van der Sluys}, M. 2004, \aap, 425, 207

\end{thebibliography}

\end{document}